\documentclass[12pt]{article}
\usepackage{amsmath}
\usepackage{graphicx}
\usepackage{pgfplots}
\usepackage[all]{nowidow}
\usepackage[utf8]{inputenc}
\usepackage{tikz}
\usepackage{multicol}
\usepackage{algpseudocode,algorithm,algorithmicx}

\newenvironment{proof}{\paragraph{Proof:}}{\hfill$\square$}

\usepackage[numbers, sort&compress]{natbib}

\usepackage{scalerel}
\usepackage{amsfonts}
\usepackage{amssymb}
\usepackage{authblk}
\usepackage{hyperref}
\usepackage[inline]{enumitem} 
\definecolor{blue}{HTML}{1F77B4}
\definecolor{orange}{HTML}{FF7F0E}
\definecolor{green}{HTML}{2CA02C}

\pgfplotsset{compat=1.14}

\usepackage{tikz}
\usetikzlibrary{arrows,shapes.arrows,shapes.geometric,shapes.multipart, decorations.pathmorphing,positioning,shapes.swigs}

\DeclareMathOperator{\E}{\mathbb{E}}
\def\spacingset#1{\renewcommand{\baselinestretch}%
{#1}\small\normalsize} \spacingset{1}

\newcommand{\eqn}{\begin{eqnarray}}
\newcommand{\ee}{\end{eqnarray}}
\newcommand{\eqnn}{\begin{eqnarray*}}
\newcommand{\een}{\end{eqnarray*}}
\newcommand{\ea}{\end{align}}
\newcommand{\be}{\begin{eqnarray}}

\newcommand{\ba}{\begin{align}}

\usepackage[english]{babel}
\newtheorem{theorem}{Theorem}
\newtheorem{lemma}[theorem]{Lemma}

\providecommand{\keywords}[1]
{
  \small	
  \textbf{\textit{Keywords---}} #1
}

\title{Marginal Structural Illness-Death Models for Semi-Competing Risks Data}

\author[1]{Yiran Zhang}
\author[3]{Andrew Ying}
\author[1]{Steve Edland}
\author[4]{Lon White}
\author[1,2]{Ronghui Xu}
\affil[1]{Biostatistics and Bioinformatics, School of Public Health and Human Longevity Science}
\affil[2]{Department of Mathematics and Halicioglu Data Science Institute, University of California, San Diego}
\affil[3]{Google Inc.}
\affil[4]{Pacific Health Research and Education Institute, Honolulu, Hawaii}
\date{}                     
\begin{document}
\maketitle
%
%
%


%
%

\begin{abstract}
The three-state illness–death model has been established as a general approach for regression analysis of semi-competing risks data. For observational data the marginal structural models (MSM) are a useful tool, under the potential outcomes framework to define and estimate parameters with causal interpretations. 
In this paper we introduce a class of marginal structural illness-death models for the analysis of observational semi-competing risks data. We consider two specific such models, the  Markov illness–death MSM and the frailty-based Markov illness–death MSM. 
For interpretation purposes, risk contrasts under the MSMs are defined.  Inference under the illness-death MSM can be carried out using estimating equations with inverse probability weighting, while inference under the frailty-based illness-death MSM requires a weighted EM algorithm. We study the inference procedures under both MSMs using extensive simulations, and apply them to the analysis of mid-life alcohol exposure on late life cognitive impairment as well as mortality using the Honolulu-Asia Aging Study data set. The R codes developed in this work have been implemented in the R package \textit{semicmprskcoxmsm} that is publicly available on CRAN.


\end{abstract}
\section{Introduction}

Our work was motivated by 
the  longitudinal epidemiologic  Honolulu-Asia Aging Study (HAAS),  
a continuation of the Honolulu Heart Program (HHP) which was a prospective, community-based cohort study of heart disease and stroke established in 1965 
with about 8,000 men of Japanese ancestry living on the island of Oahu, who were born between 1900-1919 \citep{gelber:etal:12}. HAAS was established in 1991 and 
was brought to closure in 2012 with the goal of determining the prevalence, incidence, and risk factors for Alzheimer's disease (AD) and brain aging. Demographic data, vital status and diet data were collected every 2-3 years 
during the HHP period, and neuropsychologic assessment were performed every 2-3 years 
during the HAAS. Our goal is to assess the causal effect of mid-life alcohol exposure captured during HHP on late life outcomes 
 collected in HAAS. 
In particular, 
a subject may develop 
cognitive impairment, 
then die, 
 or die without 
  cognitive impairment. 
  These are referred to as  semi-competing risks where 
  there are non-terminal events (cognitive impairment) and terminal events (death). 
  As outcomes we are interested in time to non-terminal event and time to terminal event, as well as
    time to the terminal event following the non-terminal event. 

For observational data, the potential outcomes framework of 
\citet{neyman1923application} and \citet{rubin2005causal} and the  marginal structural models    \citep[MSM]{robins2000marginal, hernan2001marginal} have been 
 a valuable tool for identifying causal effects, which can then be consistently estimated using the inverse probability of treatment weighting. 
In this paper we propose a class of marginal structural  models for the semi-competing risks data.
These structural models give rise to interpretable causal quantities including different types of risk contrasts. 
Before introducing the potential outcomes in the next section, for 
the remainder of this section we provide a review of  the three-states illness-death model that forms the basis of our structural models. 
Then in Section 3 we discuss inference under the Markov illness-death MSM and Section 4 the frailty-based Markov illness-death MSM, where a weighted EM algorithm is developed and studied. In Section 5 we carry out extensive simulation studies to assess the performance under the two MSMs including when either one of the model is valid while the other is not. We apply the approaches to the HAAS data set described above in Section 6 and conclude with more discussion in the last section.

\subsection{A review of three-state illness-death model}

\begin{figure}[!ht]
\centering
\begin{tikzpicture}[scale=0.6, transform shape]
\tikzset{line width=1.5pt, outer sep=0pt,
ell/.style={draw,fill=white, inner sep=2pt,
line width=1.5pt},
swig vsplit={gap=5pt, line color right=red,
inner line width right=0.5pt}};

\node[name=V1, ell, shape=rectangle, text width=2.5cm,text height = 0.4cm,text centered,font=\sffamily] {\Large Healthy (State 0)};
\node[name=V2, below right=30mm of V1, ell, xshift=-0.5cm, yshift=-0.5cm, shape=rectangle, text width=2.5cm,
		text centered,text height = 0.4cm,font=\sffamily]{\Large Death (State 2)};
\node[name=V3, above right=30mm of V1, xshift=-0.5cm, yshift=-0.5cm, ell, shape=rectangle, text width=2.5cm,
		text centered,text height = 0.4cm,font=\sffamily]{\Large Cognitive impairment (State 1)};

\draw[->,line width=1.5pt,>=stealth](V1) to (V2);

\draw[->,line width=1.5pt,>=stealth](V1) to (V3);

\draw[->,line width=1.5pt,>=stealth](V3) to (V2);

\end{tikzpicture}
\caption{Three-state illness-death system}\label{semicomrsk}
\end{figure}
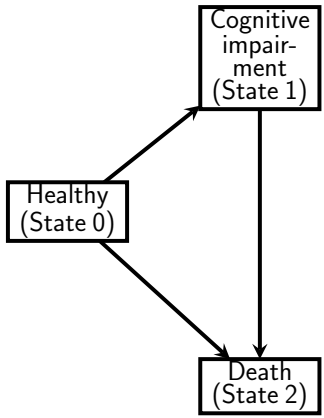

The  semi-competing risks setting introduced in the example at the beginning is the same as 
 the three-states illness-death model 
depicted in Figure \ref{semicomrsk}, 
which was first introduced by \citet{fix1951simple}. 
While  the term `model' has been used in the literature, it is more helpful to view it as a `system',  
 and not to confuse with model assumptions like Cox, Markov, etc. 
The three-state system encompasses states, transition events, time to events, and transition rates/intensities/hazards as described below. 
The system will also correspond to counterfactual outcomes later. 

We assume that 
a subject starts  in the 
 ``healthy" 
  state (state 0), 
then transition into the cognitive impairment (state 1) or death state (state 2), which are also referred to as the intermediate or non-terminal,  and the terminal state, respectively. 
The corresponding transition events are then the non-terminal event  and the terminal event, respectively.

\begin{figure}[!ht]
\centering
\includegraphics[width=80mm]{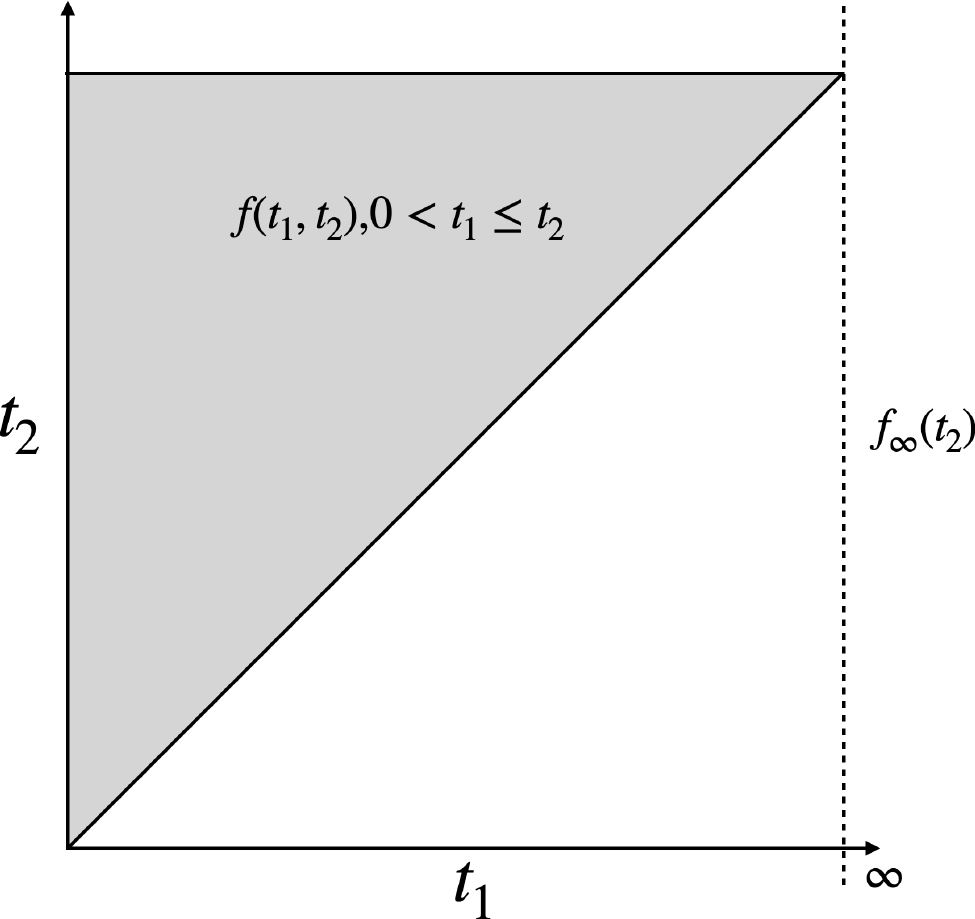}
\caption{Joint density function of  time to non-terminal event $T_1$ (eg.~cognitive impairment), and time to terminal event $T_2$ (eg.~death).} \label{jointplot}
\end{figure}

For our setup, assume a well-defined time zero, and let random variables $T_1$ and $T_2$ 
 denote time to the non-terminal and the terminal event  since  time zero, respectively. 
 If a subject 
does not experience the non-terminal event before the terminal event, we define $T_1 = +\infty$ \citep{fine2001semi, xu2010statistical}. 
Denote the joint density of $T_1$ and $T_2$ as $f(t_1,t_2)$ in the upper wedge $0<t_1 \leq t_2$, and the density of $T_2$  along the line  $t_1 = +\infty$ as $f_{\infty}(t_2)$ for $t_2 > 0$. 
Note that for semi-competing risks data, we do not observe any data in the lower wedge $0<t_2 < t_1 < +\infty$; see Figure \ref{jointplot}. We also denote the bivariate survival function of $T_1$ and $T_2$ in the upper wedge as $S(t_1,t_2)$. 

The multi-state model quantifies event rates and event risks based on the history of events,        
and is completely specified by the three transition intensities in \eqref{threestate1} - \eqref{threestate3}, also referred to as transition rates in the literature. 
Let $\lambda_1 (t_1 )$ and $\lambda_2 (t_2)$  be the transition rates from the initial healthy state to the non-terminal,  and the terminal state, respectively, 
and $\lambda_{12} (t_2 \mid t_1  )$  the transition rate from the  non-terminal state to the  terminal state.  
That is, 
\ba
\lambda_1 (t_1 )&= \lim_{\Delta \to 0^+}  \frac{  P\left(T_1 \in [t_1,t_1+\Delta) \mid T_1 \geq t_1, T_2 \geq t_1  \right)            }{      \Delta          },  \label{threestate1} \\
\lambda_2 (t_2  )&= \lim_{\Delta \to 0^+}  \frac{  P\left(T_2 \in [t_2,t_2+\Delta) \mid T_1 \geq t_2, T_2 \geq t_2  \right)            }{      \Delta          },  \label{threestate2} \\
\lambda_{12} (t_2 \mid t_1  )&= \lim_{\Delta \to 0^+}  \frac{  P\left(T_2 \in [t_2,t_2+\Delta) \mid T_1 = t_1, T_2 \geq t_2  \right)            }{      \Delta          }. \label{threestate3} 
\end{align} 
Note that \eqref{threestate1} and \eqref{threestate2} are in fact the cause-specific hazards in the usual competing risks 
 setting, for time to the non-terminal event and time to the terminal event without non-terminal event, respectively. 
In general, $\lambda_{12} \left(t_2 \mid t_1  \right)$ can depend on both $t_1$ and $t_2$. 
In the following we  consider the commonly used Markov assumption:
 $\lambda_{12} (t_2 \mid t_1 ) = \lambda_{12} (t_2)$, i.e.~the transition rate from non-terminal to terminal state does not depend on what value $T_1$ takes. 
 
 In the presence of right censoring, such as lost to follow-up or administrative censoring, let  $C$ be the time to right censoring since time zero. 
Denote $X_{1} = \min(T_{1},T_{2},C)$, $X_{2} = \min(T_{2},C)$, and the event indicators $\delta_{1} = I\left\{X_{1}=T_{1}\right\}$, $\delta_{2} = I\left\{X_{2}=T_{2}\right\}$, where $I(\cdot)$ is the indicator function. 

While the transition rates in \eqref{threestate1} - \eqref{threestate3} completely specifies the three-state illness-death model, for interpretation purposes various risk type quantities can be of interest in practice. 
Cumulative incidence function (CIF) are commonly used for competing risks \citep{kalbfleisch2011statistical};  that is, for the non-terminal event, denoted by $F_{1}(t_1)$ below, and for the terminal event without the non-terminal event, denoted by $F_{2}(t_2)$ below. In addition, we may also consider a third CIF, denoted by $F_{12}(t_1, t_2)$, for the terminal event following the non-terminal event \citep{meira2019estimation}. 
We have
\ba
& F_{1}(t_1)  = P(T_1 \leq t_1, \delta_1 = 1) = \int_0^{t_1} S(u) \lambda_{1} (u ) d u, \label{ciff1} \\ 
& F_{2}(t_2)  = P(T_2 \leq t_2, \delta_2 = 1, \delta_1 = 0) = \int_0^{t_2} S(u) \lambda_{2} (u ) d u, \label{ciff2} \\ 
& F_{12}(t_1,t_2)  =   P(T_2 \leq t_2 \mid T_1 \leq t_1, T_2  \geq  t_1) = 1 - \exp \left\{-\int_{t_1}^ {t_2}  \lambda_{12} (u) d u \right\}, \label{ciff3}
\end{align}
where $S(t) = \exp\left[- \int_0^{t} \left\{ \lambda_{1} (u) + \lambda_{2} (u ) \right\} d u \right]$.

\citet{xu2010statistical} 
discussed extensively  the illness-death model for semi-competing risks data,
and also incorporated a  shared  frailty term in  the illness-death model that encompasses previous works such as the copula model of \citet{fine2001semi}. The illness-death model with  shared frailty  has been extended to different situations 
 including in the presence of left truncation \citep{lee2021fitting}, or for a nested case-control study \citep{jazic2020estimation}.
 \citet{lee2015bayesian} extended this model to the Bayesian paradigm. 
\citet{alvares2019semicomprisks} 
developed an R package 
to 
analyze 
semi-competing risks data under the illness-death model  using parametric models and the Bayesian method, 
but not for the semiparametric Cox model formulation in the usual frequentist approach 
 to our best knowledge.

\section{Potential outcomes and structural models }

Let $A = \{0,1\}$ be a binary treatment assignment, possibly not randomized, given at or before time zero. 
Following \citet{neyman1923application} and \citet{rubin2005causal} framework of potential outcomes, we 
consider {\it counterfactual three-state} systems since time zero that consist of counterfactual states. 
That is, for different values of  $a$, the same subject may potentially transition to different states, and at different times. 
Denote $T_{1}^{a}, T_{2}^{a}, C^{a}$ as potential time to the non-terminal event, terminal event and censoring under treatment  $a=0, 1$. And $X_{1}^a$, $X_{2}^a$, $\delta_{1}^a$ and $\delta_{2}^a$ are similarly defined.

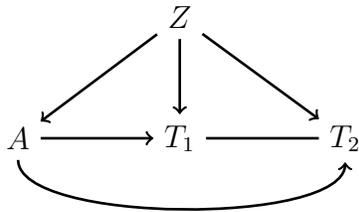
\begin{figure}[!ht]
\centering
\begin{tikzpicture}
\node[text centered] (a) {$A$};
\node[right = 1.5 of a, text centered] (t1) {$T_1$};
\node[right=1.5 of t1, text centered] (t2) {$T_2$};
\node[above = 1 of t1, text centered] (z) {$Z$};

\draw[->, line width= 1] (a) --  (t1);
\draw [-, line width= 1] (t1) -- (t2);
\draw[->, line width= 1,] (z) -- (a);
\draw[->,line width= 1] (z) --(t1);
\draw[->,line width= 1] (z) -- (t2);
\draw[->, line width=1] (a) to  [out=270,in=270, looseness=0.5]  (t2);
\end{tikzpicture}
\caption{Causal chain graph representation of semi-competing risks data, where $A$ denotes treatment, $T_1$ is time to non-terminal event (eg.~cognitive impairment), $T_2$  is time to terminal event (eg.~death), and $Z$ is a vector of covariates.}\label{chainDAG}
\end{figure}

Let Z be a $p$-dimensional vector of covariates. 
Denote $\pi(Z) = P(A=1 | Z)$, often referred to as the propensity score. 
The causal relationship of the variables defined above can be depicted in a graphical display called a chain graph  as in Figure \ref{chainDAG}, where the undirected line indicates correlation \citep{laur:rich, tchetgen2021auto}. A chain graph without undirected edges is known as a causal directed acyclic graphs (DAG).

We assume the following, which are  commonly used  in order to identify the causal estimands to be 
specified later:

(I) Stable unit treatment value assumption (SUTVA): there is only one version of the treatment and that there is no interference between subjects.

(II) Exchangeability: $(T_1^a , T_2^a, C^a) \perp A \mid Z$.

(III) Positivity: $\pi(Z) >0$. 

(IV) Consistency: If $A=a$, then $T_1^a = T_1$, $T_2^a = T_2$, 
$C^a = C$. 
\\
Exchangeability 
implies that within levels of the variable $Z$, the potential event times $(T_1^a,T_2^a)$ and the treatment assignment $A$ are independent. 
It is also called (conditional) ignobility, and that there are no unmeasured confounders. The positivity assumption 
requires that the probability of receiving either treatment ($A=1$) or control ($A=0$) 
is positive for any given value of $Z$. 
The consistency assumption here links the potential outcomes with the observed outcomes. 
For more discussion on these assumptions, please see \citet{hernan2021whatif}.

We also assume: 

(V) Non-informative censoring: $(T_1^a,T_2^a) \perp C^a \mid Z$.

Let $\lambda_1 (t_1; a )$, $\lambda_2 (t_2; a )$ and $\lambda_{12} (t_2|t_1; a )$
 be the transition rates corresponding to the counterfactual states 
 under the three-state system, $a=0, 1$. 
 As mentioned before, the transition rates completely characterize the potential three-state system; and for interpretation purposes, counterfactual risks quantities can be computed from these rates. 
\citet{andersen1991non} discussed  modeling each transition intensity by a Cox type proportional intensities regression model. Following the same idea, we can postulate the semiparametric Cox models for these transition rates, 
which are also  hazard functions \citep{andersen1991non, xu2010statistical}. In particular, we consider the following {\it Markov illness-death MSM}:
\ba
\lambda_{1} (t_1; a )&=  \lambda_{01} (t_1 ) e^{\beta_1 a}, \,\,\, t_1>0 \label{marg1}; \\
\lambda_{2} (t_2; a )&=  \lambda_{02} (t_2 )  e^{\beta_2 a}, \,\,\, t_2>0 \label{marg2}; \\
\lambda_{12} (t_2|t_1; a )&=   \lambda_{03} (t_2 ) e^{\beta_3 a }, \,\,\, 0<t_1<t_2.  \label{marg3}
\end{align}
The joint distribution  of $T^a_1$ and $T^a_2$ under model \eqref{marg1} - \eqref{marg3} will be given as a special case in the following.

The Markov illness-death model can be extended by incorporating a frailty term \citep{xu2010statistical}, to the 
{\it frailty-based Markov illness-death MSM}.  
The frailty term induces further correlation between $T^a_1$ and $T^a_2$, beyond what is already contained in the joint distribution  of $T^a_1$ and $T^a_2$.  It also  models unobserved heterogeneity among individuals \citep{lancaster1980analysis, nielsen1992counting}.
 Following \citet{vaida2000proportional} we consider the log-normal distribution for the frailty, and we have
\ba
\lambda_{1}\left(t_1| b; a \right)&=  \lambda_{01} ( t_1 ) e^{\beta_1 a+b},\,\,\, t_{1}>0 ; \label{phmm1} \\
\lambda_{2}\left(t_2| b; a  \right)&=  \lambda_{02} ( t_2 )  e^{\beta_2 a+b}, \,\,\, t_{2}>0 ; \label{phmm2} \\
\lambda_{12}\left(t_2| t_1, b; a  \right)&=   \lambda_{03} ( t_2 ) e^{\beta_3 a+b}, \,\,\, 0<t_{1}<t_{2},  \label{phmm3}
\end{align}
where $b \sim N(0,\sigma^{2})$. 
Obviously model \eqref{marg1} -  \eqref{marg3} is a special case of \eqref{phmm1} -  \eqref{phmm3} by setting $b=0$.

Recall the joint density  $f(t_1,t_2)$ and  the bivariate survival function   $S(t_1,t_2)$ 
previously defined in the upper wedge $t_1 \leq t_2$, and the density function $f_{\infty}(t_2)$ along the line  $t_1 = +\infty$. 
In the Supplementary Materials we show  that these quantities can be derived as functions of the transition rates \eqref{threestate1} - \eqref{threestate3}. 
With the models specified in \eqref{phmm1} - \eqref{phmm3} we then have the following quantities that will be used later:  
\begin{align}
f(t_{1},t_{2};a)&= \lambda_{01}(t_1) \lambda_{03}(t_2) e^{\beta_1 a + b + \beta_3 a + b} 
\exp \left\{ -\Lambda_{01}(t_1)e^{\beta_1a + b} -\Lambda_{02}(t_1)e^{\beta_1a + b}   
-\Lambda_{03}(t_1,t_2) e^{\beta_3 a + b} \right\},  \label{jointt1}  \\ 
f_{\infty}(t_{2};a)&= \lambda_{02}(t_2) e^{\beta_2 a + b}  
\exp \left\{ -\Lambda_{01}(t_2)e^{\beta_1 a + b} - \Lambda_{02}(t_2)e^{\beta_2 a + b} \right\}, 
 \label{jointt2}  \\
S(t,t ;a) &= \exp \left\{ -\Lambda_{01}(t) e^{\beta_1 a + b} - \Lambda_{02}(t) e^{\beta_2 a + b}  \right\},  \label{jointt3}
\end{align}
where $\Lambda_{0j}(t) = \int_{0}^{t}\lambda_{0j}(u)du$ for $j=1,2$, and $\Lambda_{03}(t_1,t_2) = \Lambda_{03}(t_2) - \Lambda_{03}(t_1)$ with $\Lambda_{03}(t) = \int_{0}^{t}\lambda_{03}(u)du$. 



For the rest of this section to keep notation simple, we assume that the treatment $A$ is randomized so that we can write down the relevant probabilities   for the four scenarios in the following. 
We will  then use inverse probability weighting (IPW) to create a pseudo-randomized sample. 
The proof of identification of the parameters using the observed data under the assumptions above is shown in the Supplementary Materials.  
Denote $O_{i} = (X_{1i},  X_{2i}, \delta_{1i}, \delta_{2i}, A_{i})$ the observed data for subject $i$, 
and $L_{c}$  the likelihood conditional on the random effect $b$.
We have the following four different scenarios:

\par \textbf{(i)} Non-terminal event then censored prior to terminal event: $X_{1i} = T_{1i}, X_{2i} = C_{i}, \delta_{1i} = 1, \delta_{2i} = 0$,
\begin{align*}
L_c(O_i \mid b_i )  &= \int_{X_{2i}}^{+\infty} f(X_{1i},t_{2}) d t_{2} \\
&=  \lambda_{01}(X_{1i}) e^{\beta_1 A_i + b_i} \exp \left\{
-\Lambda_{01}(X_{1i}) e^{\beta_1 A_i + b_i}  - \Lambda_{02}(X_{1i}) e^{\beta_2 A_i + b_i} 
 -\Lambda_{03}(X_{1i},X_{2i}) e^{\beta_3 A_i + b_i} \right \};
\end{align*}

\textbf{(ii)} Non-terminal event and then terminal event: $X_{1i} = T_{1i}, X_{2i} = T_{2i}, \delta_{1i} = 1, \delta_{2i} = 1$,
\begin{align*}
L_c(O_i \mid b_i ) &= f(X_{1i},X_{2i})  \\
&= \lambda_{01}(X_{1i}) \lambda_{03}(X_{2i}) e^{\beta_1 A_i + b_i + \beta_3 A_i + b_i} 
  \exp\left\{ -\Lambda_{01}(X_{1i})e^{\beta_1 A_i + b_i}-\Lambda_{02}(X_{1i})e^{\beta_1 A_i + b_i}   
  -\Lambda_{03}(X_{1i},X_{2i}) e^{\beta_3 A_i + b_i} \right\}; 
\end{align*}

\textbf{(iii)} Terminal event without non-terminal event: $X_{1i} = T_{2i}, X_{2i} = T_{2i}, \delta_{1i} = 0, \delta_{2i} = 1$,
\begin{align*}
L_c(O_i \mid b_i ) &= f_{\infty}(X_{2i})  
= \lambda_{02}(X_{2i}) e^{\beta_2 A_i + b_i}  \exp \left\{ -\Lambda_{01}(X_{2i})e^{\beta_1 A_i + b_i} - \Lambda_{02}(X_{2i})e^{\beta_2 A_i + b_i} \right\};
\end{align*}

\textbf{(iv)} Censored before any event: $X_{1i} = X_{2i} = C_{i}, \delta_{1i} = 0, \delta_{2i} = 0$,
\begin{align*}
L_c(O_i \mid b_i ) &= S(X_{1i},X_{2i}) 
=  \exp \left\{ -\Lambda_{01}(X_{1i}) e^{\beta_1 A_i + b_i} - \Lambda_{02}(X_{2i}) e^{\beta_2 A_i + b_i}  \right\}. 
\end{align*}
Combining the above four scenarios, we have
\ba
L_{c}(O_i \mid b_i ) = & 
\big\{  \lambda_{01} ( X_{1i} ) e^{\beta_1 A_i + b_i } \big\}^{ \delta_{1i} } 
\exp \left\{-\Lambda_{01}( X_{1i} ) e^{\beta_1 A_i + b_i } \right\}
\nonumber \\
&\cdot   \big\{  \lambda_{02} ( X_{2i} ) e^{\beta_2 A_i + b_i } \big\}^{   \delta_{2i}(1-\delta_{1i})   } 
\exp \left\{- \Lambda_{02}( X_{1i} ) e^{\beta_2 A_i + b_i } \right\}
\nonumber  \\
&\cdot   \big\{ \lambda_{03}( X_{2i} ) e^{ \beta_3 A_i + b_i  } \big\}^{ \delta_{2i}\delta_{1i} } 
\exp \left\{  - \Lambda_{03}(X_{1i},X_{2i}) e^{\beta_3 A_i + b_i } \right\}. 
\label{likelihood1}
\end{align}


\section{Illness-death MSM} \label{usualM}


In the absence of randomization, denote $w_i = A_i/ \hat\pi(Z) + (1-A_i)/\{1-\hat\pi(Z)\}$
as the IP weight for subject $i$. 
It can be shown in general that under Assumptions (I)-(IV), IPW identifies the causal effects 
under marginal structural models \citep{hernan2021whatif}. 
In practice, $\pi(\cdot)$ is unknown and  can be estimated  from the data by either specifying a 
parametric model such as the logistic regression \citep{robins2000marginal},  or 
use  nonparametric methods such as boosted trees \citep{mccaffrey2004propensity}.

For the Markov illness-death MSM \eqref{marg1} - \eqref{marg3}, with $b_i=0$ in  \eqref{likelihood1} we have  the weighted log-likelihood 
\ba
\log L_w  = 
& \sum_{i} w_{i} \bigg[ \delta_{1i} \big\{ 
\beta_{1} A_{i} + \log \left(\lambda_{01}(X_{1i})\right)\big\} -  \Lambda_{01}(X_{1i})e^{\beta_{1}A_{i} 
} \bigg] \nonumber \\ 
&+ \sum_{i} w_{i} \bigg[ \delta_{2i}(1-\delta_{1i}) \big\{ 
\beta_{2} A_{i} + \log \left(\lambda_{02}(X_{2i})\right)\big\} -  \Lambda_{02}(X_{1i})e^{\beta_{2}A_{i} 
} \bigg]   \nonumber \\
&+ \sum_{i} w_{i} \bigg[ \delta_{2i}\delta_{1i} \big\{ 
\beta_{3} A_i + \log \left(\lambda_{03}(X_{2i})\right)\big\}- \Lambda_{03}(X_{1i},X_{2i})e^{\beta_{3}A_{i} 
} \bigg]. 
\label{usuallikelihoods}
\end{align}
It can be seen that the parameters for the three transition rates $(\beta_{j}, \Lambda_{0j})$,  $j=1,2,3$,
 are variationally independent in the above likelihood and therefore can be estimated separately. Note that the semiparametric approach under the Cox type models discretizes the baselines hazards $\lambda_{0j}(\cdot)$ into point masses at the observed event times and estimates the cumulative $\Lambda_{0j}(\cdot)$ as step functions.
It can  be verified that maximizing \eqref{usuallikelihoods} is equivalent to maximizing the following three {weighted} Cox regression model likelihoods: 
1) 
treating the non-terminal event as the event of interest, and terminal event without non-terminal or originally censored as ‘censored’; 
2) 
treating the terminal event without non-terminal as the event of interest, and non-terminal event or originally censored as ‘censored’; 
3) 
treating the terminal event following the non-terminal as the event of interest, left truncated at the time of the non-terminal event (so only those who had the non-terminal event are included), and originally censored as ‘censored’. 
Then the standard software (e.g.~\textit{coxph()} in R package `survival') can be used to obtain the estimates $(\hat{\beta}_j, \hat\Lambda_{0j}), j=1,2,3$.  

In order to obtain the variance of the estimates, if we assume the estimated weights in \eqref{usuallikelihoods} as known, 
then the robust sandwich variance estimator 
in standard software such as \textit{coxph()}  can  be used to obtain separately each of the estimated variance for $\hat{\beta}_j, j=1,2,3$. 
In the Supplementary Materials 
we provide  the formulas for estimating  the covariances between $\hat\beta_j, j=1,2,3$. 
In addition, we may also use the bootstrap  variance estimator which accounts for the uncertainty in estimating the weights.

For causal interpretation, we may define the risk contrasts as the difference or the ratio between the CIF's under the  structural models with $a=1$ and $a=0$. 
In particular, 
 \ba
& F_{1}(t_1; a)   = \exp(\beta_1a) \int_0^{t_1} S(u;a) \lambda_{01} (u)  d u,  \\ 
& F_{2}(t_2; a)   = \exp(\beta_2a) \int_0^{t_2} S(u;a) \lambda_{02} (u )  d u,  \\ 
& F_{12}(t_1,t_2; a)   = 1 -  \exp\left\{- e^{\beta_3a} \int_{t_1}^ {t_2}  \lambda_{03} (u)  d u \right\},  
\end{align}
where $S(t;a) = \exp\left[- \int_0^{t} \left\{ \lambda_{01} (u)e^{\beta_1a} + \lambda_{02} (u )e^{\beta_2a} \right\}  d u \right]$. 
We estimate the contrasts by plugging in the parameter estimates, and obtain their 95\% confidence intervals (CI) using bootstrap. 
We note that for simple competing risk data under the marginal structural Cox model, such risk contrasts are available in the R package `cmprskcoxmsm' \citep{cmprskcoxmsm}.

\section{Frailty-based illness-death MSM}

Under the frailty-based  Markov illness-death MSM \eqref{phmm1} - \eqref{phmm3} where
 $b \sim N(0, \sigma^2)$, let $\theta = (\beta_{1},\beta_{2},\beta_{3}, \Lambda_{01},\Lambda_{02},\Lambda_{03},\sigma^{2})$. 
Denote $O = \{O_i\}_{i=1}^n$. 
 The weighted  observed data likelihood 
  is:
\be
L_w(  \theta; O) &=&  \displaystyle \prod_{i} \big\{ \scaleobj{.7} {\int}  L_c(  \theta; O_{i} \mid b_{i}) 
\cdot f(  \theta; b_{i}) db_{i}  \big\}^{w_{i}}, \label{wfull}
\ee
where $ f(  \theta; b_{i}) $ is the normal density function.
Then the estimate $\hat{\theta}$
can be obtained by maximizing \eqref{wfull}.

We introduce below an EM type algorithm 
in order to maximize \eqref{wfull}.
Denote $ Q(  \theta,\tilde{  \theta}) $ the expectation of the weighted log-likelihood of the augmented data $\left(y_{i},b_{i}\right)$, $i=1, ..., n$, conditional on the observed data and the current parameter value $\tilde{  \theta}$:
\ba
Q(  \theta,\tilde{  \theta}) &
= \sum_{i} \E \big[w_{i} \cdot l\left(  \theta_{i}; O_{i} | b_{i} \right)   \mid  O,  \tilde{  \theta}  \big] +  \sum_{i} \E \big[ w_{i} \cdot \log f \left(  \theta; b_{i} \right)   \mid  O, \tilde{  \theta} \big], 
\end{align}
where 
\ba
l\left( \theta ;  O \mid b \right) = & 
\bigg[ \delta_{1} \left\{b + \beta_{1} A + \log \left(\lambda_{01}(X_{1})\right)\right\} \nonumber \\
&+ \delta_{2}(1-\delta_{1}) \left\{b + \beta_{2} A + \log \left(\lambda_{02}(X_{2})\right)\right\}+ \delta_{2}\delta_{1} \left\{b + \beta_{3} A + \log \left(\lambda_{03}(X_{2})\right)\right\} \nonumber \\
&- \Lambda_{01}(X_{1})e^{\beta_{1}A+b} - \Lambda_{02}(X_{1})e^{\beta_{2}A+b} - \Lambda_{03}(X_{1},X_{2})e^{\beta_{3}A+b}\bigg]. \label{wcond}
\end{align}
Then $ Q = Q_1 + Q_2 + Q_3 + Q_4$, where 
\ba
Q_{1}(\beta_{1},\lambda_{01}) &= \sum_{i} w_{i} \bigg[ \delta_{1i} \big\{ \E(b_{i}) + \beta_{1} A_{i} + \log \left(\lambda_{01}(X_{1i})\right) \big\} -  \Lambda_{01}(X_{1i})
\exp\{\beta_{1}A_{i}+ \log \E(e^{b_i})\} \bigg], \label{Q1} \\ 
Q_{2}(\beta_{2},\lambda_{02}) &= \sum_{i} w_{i} \bigg[ \delta_{2i}(1-\delta_{1i}) \big\{\E(b_{i}) + \beta_{2} A_{i} + \log \left(\lambda_{02}(X_{2i})\right)\big\} -  \Lambda_{02}(X_{1i})
\exp\{\beta_{2}A_{i}+\log \E(e^{b_i}) \} \bigg],   \label{Q2} \\
Q_{3}(\beta_{3},\lambda_{03}) &= \sum_{i} w_{i} \bigg[ \delta_{2i}\delta_{1i} \big\{\E(b_{i})+ \beta_{3} A_i + \log \left(\lambda_{03}(X_{2i})\right)\big\}- \Lambda_{03}(X_{1i},X_{2i})
\exp\{\beta_{3}A_{i} +\log \E(e^{b_i}) \} \bigg],  \label{Q3}  \\
 Q_{4}(\sigma^{2}) &= \sum_{i} w_{i} \bigg\{ -\frac{1}{2} \big(\log 2\pi + \log \sigma^{2} \big)-\frac{1}{2\sigma^{2}} \E(b_{i}^{2}) \bigg\},  \label{Q4}
\end{align}
where $E\{h(b_i)\} = E\{h(b_i) \mid O_i,\tilde{  \theta} \}$ is shorthand for a function $h(\cdot)$ of $b_i$. 
Analogous to the EM  algorithm, 
we iterate  between the E-steps and the M-steps described below until convergence. 

\subsection*{E-step}

The 
conditional expectations in \eqref{Q1} -  \eqref{Q4} 
are all in form of $E\{h(b_i) \mid O_i,\tilde{  \theta} \} = \int h(b_i) f(b_i \mid O_i,\tilde{  \theta}) db_i$, where $h(b_i) = e^{b_i} $ in \eqref{Q1} -  \eqref{Q3} and $h(b_i) = b_{i}^{2} $ in \eqref{Q4}. 
These two expectations are not in closed form; however, we can approximate these integrals by numerical methods, specifically by (adaptive) Gaussian quadrature \citep{gander2000adaptive, rice1975metalgorithm}. 
Details of computation are shown in the Supplement Materials. 

\subsection*{M-step}

The M-step conveniently separates the update of  $\beta_{j}$ and $\Lambda_{0j}$ for $j=1,2,3$ from that of the variance component $\sigma^{2}$. 
For $Q_1$ - $Q_3$, similar to Section \ref{usualM},  \eqref{Q1} - \eqref{Q3} are equivalent to
the weighted  log-likelihood functions in a Cox regression  with additional known offsets $ \mu_{i} = \log E(e^{b_i} \mid O, \tilde{  \theta})$. 
In order to maximize $Q_{4}$, we set 
\ba 
\frac{\partial Q_{4}}{\partial \sigma^{2}} &= \sum_{i} w_{i} \bigg\{-\frac{1}{2\sigma^{2}} + \frac{\E(b_{i}^2 \mid  O,  \tilde{  \theta})}{2\sigma^{4}} \bigg\} = 0, \nonumber
\end{align}
leading to 
\ba 
\hat\sigma^{2} = \frac{\sum_{i=1}^{n} w_{i} \E(b_{i}^2 \mid  O,  \tilde{  \theta})}{\sum_{i=1}^{n} w_{i} },
\end{align}

In the  lemma below, 
we establish the following property of the above weighted EM algorithm, which is similar to that of the EM algorithm. 
\begin{lemma}\label{lemma1} 
Suppose $L_{w}(  \theta; O)$  is the weighted observed data likelihood. 
At step $k$ of  the  algorithm denote $  \theta^{(k)} $ the current value, and $ \theta^{(k+1)} $ the value that maximizes  $Q(  \theta, \theta^{(k)} )$.
Then:
\be
L_{w}(  \theta^{(k+1)};O) \ge L_{w}(  \theta^{(k)};O).
\ee
\end{lemma}
The proof of the lemma is given in the Supplement Materials. 
Following \citet{wu1983convergence} 
or Theorem 4.12 
 in \citet{lehmann2006theory}, since 
$Q(\theta ; \tilde\theta)$ is continuous in both $\theta$ and $\tilde\theta$, then all limit points of the weighted EM sequence $\{ \theta^{(k)} \}$ are stationary points of $L_{w}(  \theta; O)$, and $L_{w}(  \theta^{(k)}; O)$ converges monotonically to $L_{w}(  \theta^*; O)$ for some stationary point $\theta^*$. 
In addition, for existence of such limit point(s) \citet{vaida2005parameter} proposed a condition for the usual unweighted EM algorithm: 
as long as the maximizer in the M-step is unique. 
We can show that this result extends immediately to our weighted EM algorithm.
And finally,  our M-step satisfies this condition, i.e.~the maximizer in the M-step is unique. 

As initial values 
we use  for $\beta_j$ and $\Lambda_{0j}$, $j=1,2,3$, the estimates from weighted Cox regression without the offsets, i.e.~from the illness-death MSM of the previous section;  
and  $\sigma^2 =1$. 
The stop criteria we use in this paper are convergence 
in the log-likelihood 
as well as in parameters of interest: $| \log L_{w}(  \theta^{(k+1)};y) - \log L_{w}(  \theta^{(k)};y) | \le 10^{-5}$, $ | \beta_j^{(k+1)} - \beta_j^{(k)} | \le 10^{-3} $, $j=1,2,3$ and $| {\sigma^2}^{(k+1)} - {\sigma^2}^{(k)} | \le 10^{-3}$.

\subsection*{Variance estimate}

The variance of the parameter estimates following a
typical  EM algorithm can be estimated by the inverse of a (discrete) observed information matrix calculated using Louis’ formula, including for the nonparametric maximum likelihood estimator
 (NPMLE) under, for example, the semiparametric proportional hazards mixed models \citep{vaida2000proportional}. 
For observational data, however, inference using the weighted NPMLE under semiparametric models requires the derivation of efficient influence functions \citep{breslow2007weighted},  
 and is generally non-trivial under the normal frailty construct \citep{murp:vand:00, mapl:murp:02}. 
In the following  we use bootstrap to obtain the variance estimator for $\hat{\theta}$. 

\subsection*{Risk contrasts}

Similar to what we proposed under the illness-death MSM, we also can define the risk contrasts under the frailty-based illness-death MSM. Since the frailty-based models are conditional on the random effect $b$, we have the following conditional risk:
\ba
& F_{1}(t_1 \mid b ; a)   = \exp(\beta_1a + b) \int_0^{t_1} S(u \mid b ;a) \lambda_{01} (u)  d u \label{generalcifb1},  \\ 
& F_{2}(t_2 \mid b ; a)   = \exp(\beta_2a + b) \int_0^{t_2} S(u \mid b ;a) \lambda_{02} (u )  d u,  \label{generalcifb2} \\ 
& F_{12}(t_1,t_2 \mid b ; a)   = 1 -  \exp \left\{- e^ {\beta_3a + b} \int_{t_1}^ {t_2}  \lambda_{03} (u)  d u \right\},  \label{generalcifb3}
\end{align}
where $S(t \mid b;a) = \exp\left[- \int_0^{t} \left\{ \lambda_{01} (u)e^{\beta_1a + b} + \lambda_{02} (u )e^{\beta_2a + b} \right\} d u \right] = \exp\left\{- e^{\beta_1a + b} \Lambda_{01}(t) - e^{\beta_2a + b}  \Lambda_{02} (t ) \right\}$. 

As discussed earlier the frailty term, or equivalently, the random effect $b$ represents the unobserved heterogeneity among the individuals, which is assumed be unaffected by the treatment $a$ received. As such, the above conditional risk represents individual risk, and the risk contrasts the individual risk contrasts. We therefore have  the individual risk difference (IRD) and the individual risk ratio (IRR). 
Under the random effects model, for  $ i=1,2,...,n$, the predicted random effect is $\hat{b}_i = \E(b_i \mid O_i, \hat{\theta})$ \citep{vaida2000proportional}.  We then obtain the predicted IRD and the predicted IRR. 
For inference on these individual risk contrasts, Bayesian bootstrap \citep{kosorok2008introduction} may be used which, unlike the usual resampling with replacement, preserves each individual $i$ in the original data set. Details of the Bayesian bootstrap are provided in the Supplementary Materials. 
Note that because $b$ is random, the common terminology in the literature is `predicted' instead of `estimated', and `prediction interval (PI)' instead of CI. 


\section{Simulation}

We carry out extensive Monte Carlo simulation studies in order to assess the performance of the 
estimation procedure described above.  
We use the idea  from \citet{havercroft2012simulating} to simulate  data under the marginal structural model \eqref{phmm1} - \eqref{phmm3}. We also adapt the  method from \citet{jiang2015simulation}, 
originally designed for simulating semi-competing risk data with gamma frailty.
Very briefly the following steps are used to to generate the data; more details are provided in the Supplementary Materials. 
\begin{itemize}
  \item Generate $U_{1} \sim U(0,1)$ and $U_{2} \sim U(0,1)$;
  \item Generate confounder $Z = (Z_{1}, Z_{2}, Z_{3})^{\intercal}$, with $Z_{j} = U_{1} + U_{2} + \epsilon_{j}, j = 1,2,3$, where $\epsilon_{1} \sim N(0,1)$,  $\epsilon_{2} \sim N(0,1.5)$ and $\epsilon_{3} \sim N(0,1.8)$;
  \item Generate $A \sim$ Bernoulli($p_{A}$), where $p_{A} = \mbox{logit}^{-1}(\alpha_0 + \alpha_1 Z_1 + \alpha_2 Z_2 + \alpha_3 Z_3)$, with $\alpha_0 = 0.5, \alpha_1 = 0.1, \alpha_2 = -0.1, \alpha_3 = -0.2$;
  \item Let $\lambda_{01}\left(t \right) = \lambda_{02}\left(t \right) = 2e^{-t}I(0 \le t \le 3) + 2e^{-3}I(t > 3)$ and $\lambda_{03}\left(t \right) = 2 \lambda_{01}\left(t \right )$. Then with probability $P(T_{1}=\infty)$ 
  given  in the Supplementary Materials, 
  $$T_{2} = \Lambda_{01}^{-1}\left(-\frac{\log(U_{1})}{\exp(\beta_{1}A+b) + \exp(\beta_{2}A+b)}\right);$$
   and with probability $1-P(T_{1}=\infty)$, 
   $$T_{1} = \Lambda_{01}^{-1}\left(-\frac{\log(U_{1})}{\exp(\beta_{1}A+b) + \exp(\beta_{2}A+b)} \right), \,\,\, T_{2} = \Lambda_{01}^{-1}\left( -\frac{\log(U_{2})}{2\exp(\beta_{3}A+b)} + \Lambda_{01}(t_{1}) \right).$$
  \item Generate Censoring time $C \sim U(0.4, 0.5)$, which  leads to an average censoring rate around 20\%.
\end{itemize}

We set $\beta_{1}=\beta_{2}=1$, $\beta_{3} = 0.5$.
 Weights are calculated by fitting the logistic regression with $Z_1, Z_2, Z_3$ as covariates. 
 We run 500 simulations for each case. 
 Table \ref{simulation_1} and \ref{simulation_2} report, for sample size n=250 and n=500, respectively, the estimate, the empirical standard deviation (SD), the mean of estimated standard errors (SE), and the coverage probability (CP) of the nominal 95\% confidence intervals.
  Under the illness-death MSM, we 
  estimate the asymptotical variance 
  of $\beta_j$, $j=1,2,3$ using both the model-based formulas, which ignores the uncertainty in the estimation of the weights,  and bootstrap.
 
When $\sigma^2 = 0$, 
we 
see that the estimation under the illness-death MSM is nearly unbiased as expected, in particular for the 
larger sample size $n=500$, and the coverage of the confidence intervals (CI) based on the normal approximation is very close to the nominal level. 
We note that the margin of error  using 500 simulation runs to estimate the coverage of 95\% CI's is 0.019, so that the range of coverage probability (CP) should be mostly within 93.1\% to 96.9\%.
We also see that when $\sigma^2 = 0$, the estimation under the frailty-based illness-death MSM performed well for $\beta_j$ and $\Lambda_{0j}(01)$, $j=1,2,3$. However, the mean of the estimated standard error of $\sigma^2$ is much higher than the empirical standard deviation, and the CI overcovers. We note that this is the boundary cases considered in  \citet{xu2009using}, where  
 the asymptotical distribution is no longer normal. 

When $\sigma^2 > 0$, 
 we 
 see that our 
 estimator under the frailty-based illness-death MSM is 
 quite accurate for even the smaller 
 sample size $n=250$, 
 the SEs are close to the sample SD and the coverage probabilities are good.
The estimates under the illness-death MSM without frailty is obviously biased, as expected, with poor coverage of the CI's when $\sigma^2 > 0$.

Finally, we note that the variances of the estimators are generally larger under the frailty-based illness-death MSM, as one more parameter is estimated.

\section{Application to HAAS study}

For this analysis, we are interested in 
 the effect of mid-life alcohol exposure on 
cognitive impairment as well as death, which are semi-competing risks. 
In the HHP-HAAS study, alcohol consumption was assessed by self-report and translated into units of drinks per month. Estimates of the total ethanol intake from reported drinking patterns were calculated as ounces per month for beer, liquor, wine, and sake using algorithms based on average unit sizes and usual alcohol percentages. 
 The alcohol consumption was then dichotomized into 
 heavy drinking ($>$2 drinks/day) or not. 
The 
``mid-life" alcohol exposure was  collected during the HHP study between 1965-73. 
The heavy drinking group consisted of individuals who had heavy drinking at one point during mid-life, and the 
other group those who never had heavy drinking  during mid-life. 
Cognitive impairment 
was based on 
scores from the Cognitive Assessment and Screening Instrument (CASI), where a score below 74 was considered a moderate impairment (MI). 

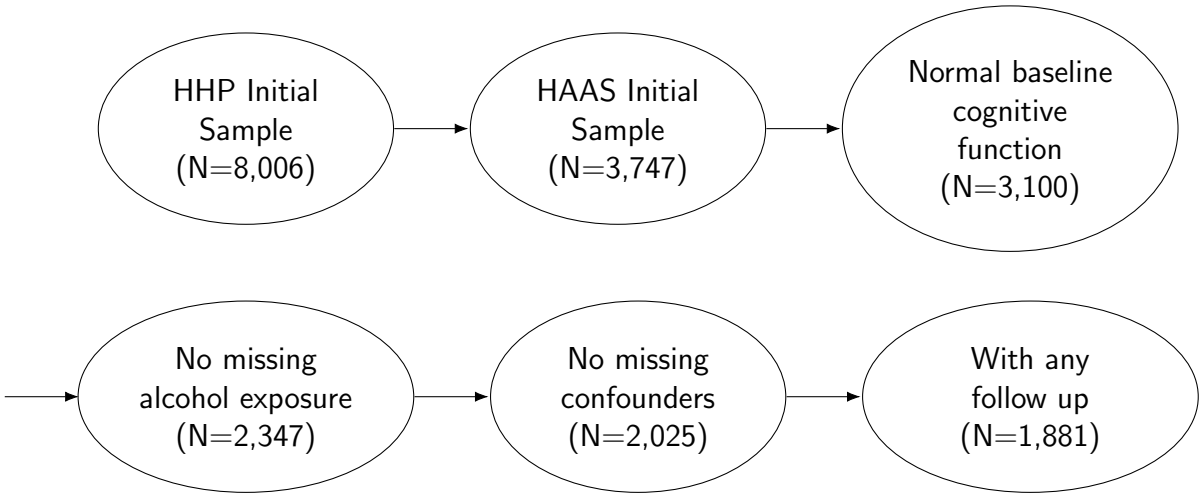
\begin{figure}[!ht]
\centering
\usetikzlibrary{shapes,decorations,arrows,calc,arrows.meta,fit,positioning}

\begin{tikzpicture}[scale=1.0, transform shape]

\tikzset{
    -Latex,
    auto,
    node distance=1cm and 1cm,
    semithick,
    state/.style={ellipse, draw, minimum width=0.7cm},
    point/.style={circle, draw, inner sep=0.04cm, fill, node contents={}},
    bidirected/.style={Latex-Latex, dashed},
    el/.style={inner sep=2pt, align=left, sloped}
};

\node[state] (v1) at (0,0) [text width=2.5cm, text height=0.4cm, text centered, font=\sffamily] {HHP Initial Sample\\(N=8,006)};
\node[state] (v2) [right=of v1, text width=2.5cm, text height=0.4cm, text centered, font=\sffamily] {HAAS Initial Sample\\(N=3,747)};
\node[state] (v3) [right=of v2, text width=2.88cm, text height=0.4cm, text centered, font=\sffamily] {Normal baseline cognitive function\\(N=3,100)};

\node[state] (v4) [below=of v1, text width=2.88cm, text height=0.4cm, text centered, font=\sffamily] {No missing alcohol exposure\\(N=2,347)};
\node[state] (v5) [right=of v4, text width=2.5cm, text height=0.4cm, text centered, font=\sffamily] {No missing confounders\\(N=2,025)};
\node[state] (v6) [right=of v5, text width=2.88cm, text height=0.4cm, text centered, font=\sffamily] {With any follow up\\(N=1,881)};

\path (v1) edge (v2);
\path (v2) edge (v3);
\path (v4) edge (v5);
\path (v5) edge (v6);

\draw[-Latex]([xshift = -155]v4.east) -- ++(1,0);

\end{tikzpicture}
\caption{Flowchart of samples since the start of HHP}\label{flowchart}
\end{figure}

The confounders were decided by literature review 
 and clinical experiences, as well as availability of the data. Literatures show that vital data such as blood pressure and heart rate are associated with drinking habits, as well as the cognitive health. Meanwhile, demographic data such as age, years of education,
 are also related to cognitive impairment and drinking habits. The Apolipoprotein E is the first identified genetic susceptibility factor for sporadic AD. Towards understanding determinants of cognitive impairment and factors associated with drinking habits, the final set of baseline confounders 
 are baseline CASI score,  systolic blood pressure, heart rate, Apolipoprotein E genotype positive, years of education and baseline age. 
We only include participants with normal cognitive function (CASI $\ge$ 74) at baseline, 
 and after excluding missing values for exposure and confounders, we have 1881 participants in total that have had any follow up since the start of HAAS. 
Figure \ref{flowchart} provides a flowchart of the samples since the start of HHP.

Since HAAS is a long-term epidemiology study, lost to follow-up occurs at every exam visit. 
On the other hand, death certificates 
were obtained for many participants, even after lost to follow-up. 
For this reason, we needed to  properly define the death 
 for the semi-competing risks data. If the death date 
  is after the participant's recorded last visit date from the study, we consider this participant lost to follow-up. 
  More details of data pre-processing can be found in \citet{yiranthesis}. 

Propensity scores (PS) were calculated using R package `twang' (Toolkit for Weighting and Analysis of Nonequivalent Groups), which estimates the PS using boosted regression as the predicted probability of being heavy drinker versus not, conditional 
on the measured baseline confounders. 
In Supplementary Materials  
we show the PS histograms in the heavy and not heavy drinking groups. 
Before applying the IPW approach to the multi-state model, we obtained stabilized weights and trimmed them within (0.1, 10).
We also plot in the Supplementary Materials the  standardized mean difference (SMD) to check the balance of each confounder before and after weighting, 
where the SMD's of all the confounders are 
within the  interval [-0.1, 0.1] after weighting.  

We apply our proposed methods to the HAAS data. 
For the purposes of this analysis, time zero is defined as the start of HAAS, and only subjects who were alive and had normal cognitive functions at time zero are eligible for the analysis. See the Discussion Section for other considerations such as using age as the time scale for analysis. 
We first fit the illness-death MSM and the results are in the top half of Table \ref{HAAS1}. 
We see that the transition rates to moderate impairment or death without moderate impairment are significantly higher with heavy drinking  compared to not heavy drinking.
 But we don't see a significant difference in 
 the transition rates to death after moderate impairment.  

We then fit the 
frailty-based illness-death MSM 
and the results are in the bottom half of Table \ref{HAAS1}. 
The convergence plot of the parameters and the likelihood during the weighted EM algorithm are provided in the Supplement Materials,  
where we stopped at 168 EM steps for the final results.
Compared to the results under the illness-death MSM without frailty, the magnitude of all three estimated effects are further away from the null, and all three transition rates are significantly higher under heavy drinking  than  not heavy drinking. 
The phenomenon of more significant and away-from-the-null exposure effects after accounting for the frailty is known in the literature under the Cox model \citep{chastang1988quantitative}.  

\begin{figure}[!h]
\centering 
\includegraphics[width=160mm]{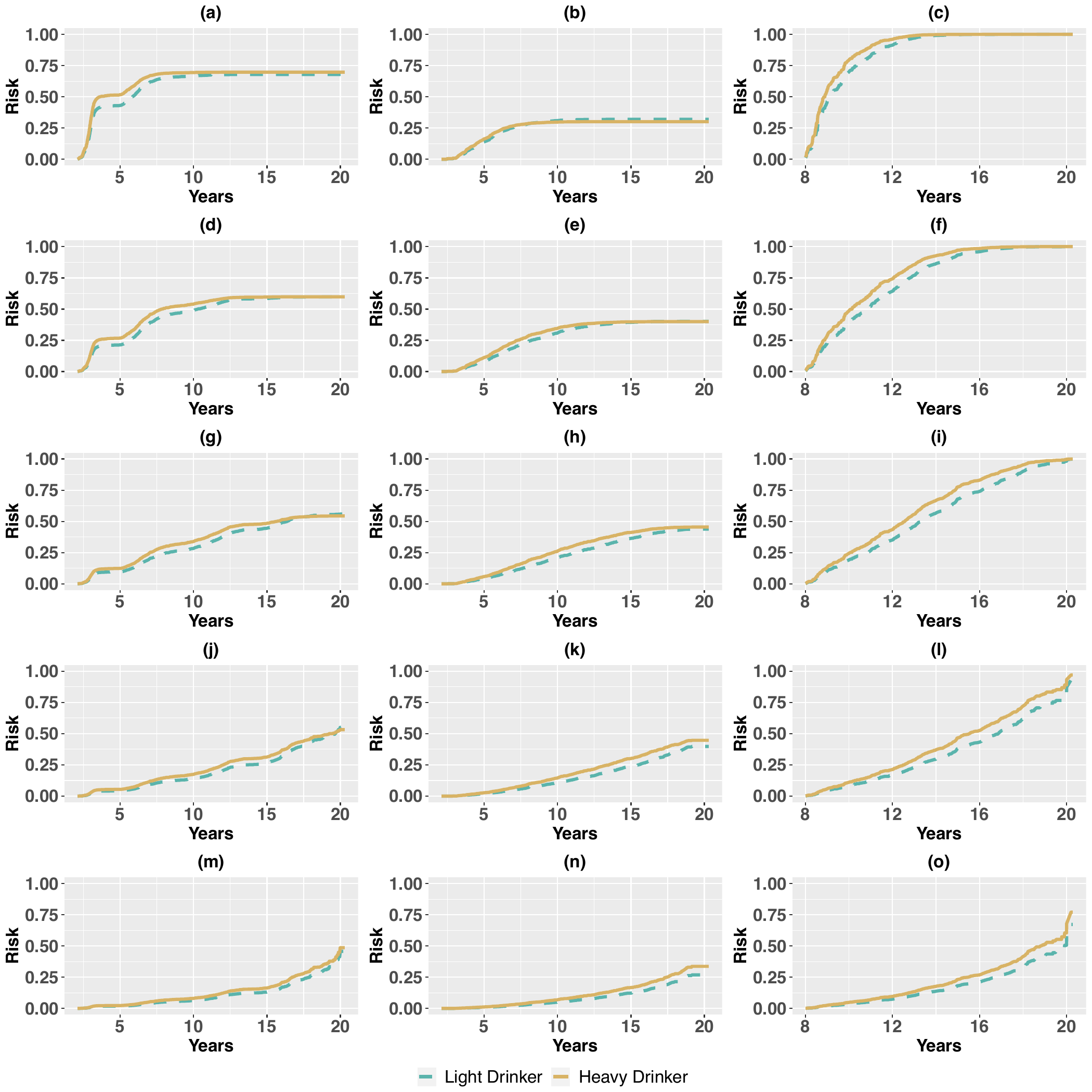}
\caption{ 
Risk plots for HAAS data under the illness-death MSM, row 1;  and conditional risk plots under the frailty-based illness-death MSM, rows 2-6, for $b =  2\hat{\sigma} $,  $ \hat{\sigma}  $,  0, $  -\hat{\sigma}$ and $   -2\hat{\sigma} $, respectively.  
The columns from left to right are:  moderate impairment (MI), death without MI, and death following MI by $t_1 = 8$ years. 
} \label{cifall}
\end{figure}

Finally, we estimate the causal risk contrasts under the structural models. 
For illustration purposes we  fix $t_1 = 8$ years in $F_{12}(t_1,t_2; a)$ and $F_{12}(t_1,t_2|b; a)$; that is, the cumulative incidence rate of death following MI by 8 years. We show the estimated risk curves in Figure \ref{cifall} first row under the illness-death MSM,  and the risk contrasts in Table \ref{HAAS2} for heavy drinking versus not. 
It is seen 
that the risk contrasts for  MI 
are significantly different from the null at 5, 10 and 15 years, but not so at  20 years. 
The risk contrasts for death without MI and death following MI by 8 years are not significantly different from the null at  any of the years under the illness-death MSM.

\begin{figure}[!h]
\centering 
\includegraphics[width=160mm]{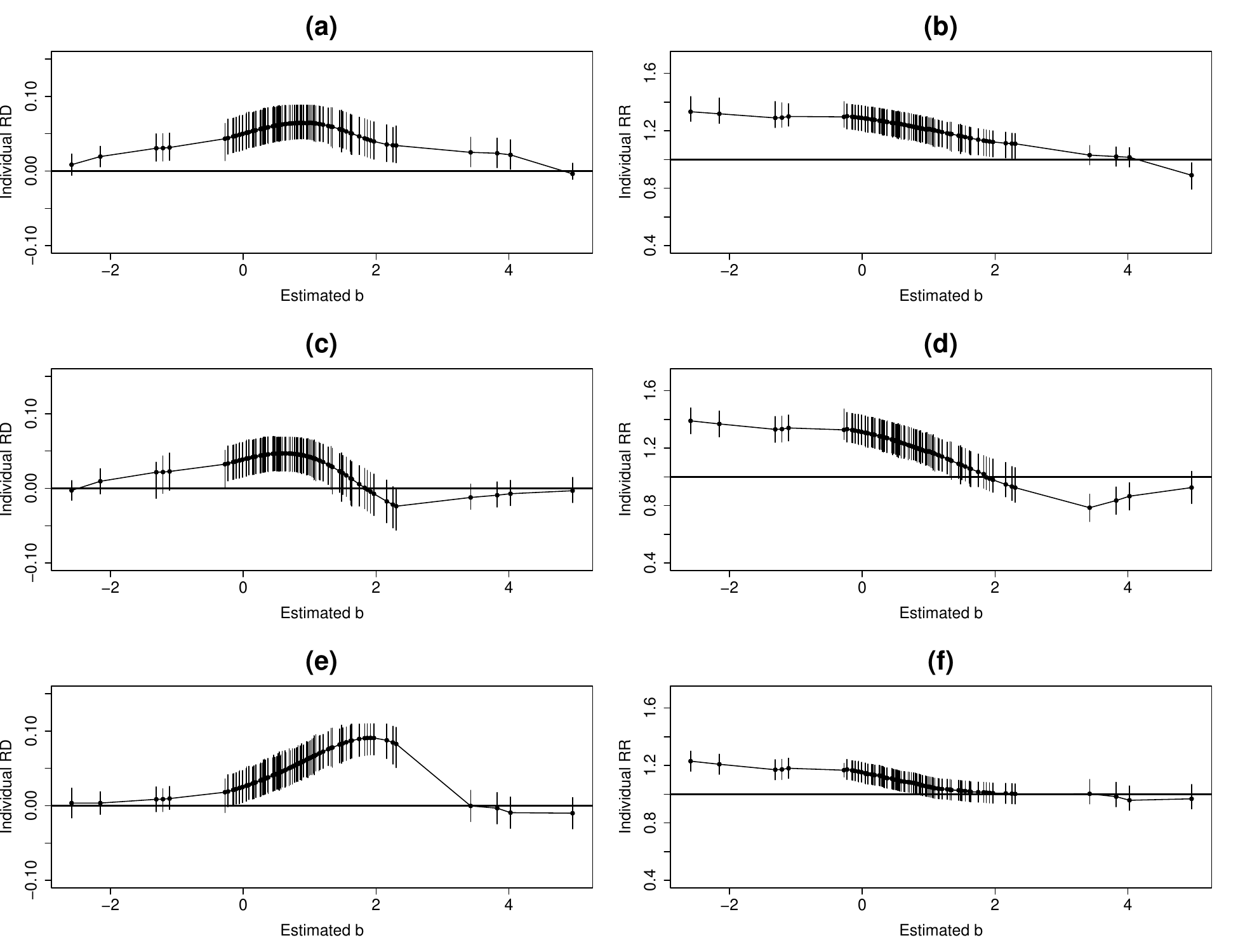}
\caption{ 
Individual risk difference (left) and individual risk ratio (right)  at 10 years with 95\% prediction intervals, for 100 participants of the HAAS study at  every percentile of the predicted $b$'s.
Top row: moderate impairment (MI); middle row: death without MI; bottom row: death following MI by $t_1 = 8$ years. } \label{IRR10}
\end{figure}

We also show the predicted conditional risk curves at different $b$ values ($0, \pm \hat{\sigma}, \pm 2\hat{\sigma}$) under the frailty-based illness-death MSM in Figure \ref{cifall}, plots (d) - (r).   
In Figure \ref{IRR10} we plot the IRD and IRR at 10 years with 95\% PI's of 100 participants from every  percentile of the predicted $b$ values. 
We note the different significance results for IRD and IRR: the IRD tends to be significantly different from the null for $b$ values closer to zero, while the IRR tends to be significantly different from the null for negative $b$ values. This appears to be generally the case for all three outcomes: MI, death without MI, and death following MI by 8 years. 
More discussion will follow in the next section.

\section{Discussion}

In this paper we proposed marginal structural models based on the three-state illness-death system for  observational semi-competing risks data using the potential outcomes framework. 
Our development parallels the original MSMs for longitudinal and survival data 
\citep{robins2000marginal, hernan2001marginal}.  Inverse probability of treatment weighting is used to fit these structural models. 
We developed estimation algorithms under both the  
illness-death MSM in the absence of frailty, and  under the frailty-based illness-death MSM  where a weighted EM algorithm is developed and its convergence property studied. 
The methods are implemented in a publicly available R package.   
The simulation studies showed the good performance of the proposed methods. 

For applications in practice, we have defined cumulative risk based causal contrasts and illustrated their use. 
Under the  frailty-based illness-death MSM, these give rise to individual risk contrasts IRD and IRR. This is consistent with the random effects modeling formulation, where individual trajectories, for example, from longitudinal data can be estimated and predicted. We have extended this feature to the causal inference setting, when the individual heterogeneity is modeled using random effects. 
It might also be of some interest to compare the IRD and IRR to the RD and RR under the illness-death MSM without frailty, and note some similarity between the first and the fourth row of Figure \ref{cifall}, where the random effect $b$ is set to its mean value of zero. We note that these two sets of contrasts are not the same, especially since the Cox model is not collapsible; and the interpretations are different for these two sets of contrasts. 

For the HAAS data other times scales might be of interest, such as age. As is typical with those types of aging study analyses, left truncation arises since subjects were at different ages at the start of HAAS, and those who have had early events (death or impairment) compared to their ages at the start would have been truncated out. Under the MSM's where exposure status is the only regressor, any covariate dependent truncation would be considered informative and need to be further handled by, for example, inverse probability of truncation weighting. This is investigated  in \cite{wang:etal:arxiv} for a single survival outcome of the HAAS data, and is being extended to more complex settings such as potential outcomes under separate projects. 

We have made the Markov assumption $\lambda_{12}(t_2|t_1)=  \lambda_{12}(t_2)$ following \cite{xu2010statistical}. This assumption may be tested by including $T_1$ as a covariate in fitting model \eqref{marg3} \citep{geskus2016data}. For the HAAS data this test gives a $p$-value of 0.83, hence the Markov assumption is not rejected.  Conceivably in this relatively old population with a maximum of about 20 years of follow-up, the event times are not as varied as studies with much longer follow-up, where the  Markov assumption might be more likely violated.

In the HAAS data of the 887 subjects with moderate impairment, 137 (15.6\%) had their CASI scores back to  normal after first dropping under the threshold of 74 points, and 72 out of those 137 subjects became impaired again subsequently. Our analysis approached used so-called first crossing time \citep{dono:etal:2011}. Alternatively, 
\cite{pak:etal:2023} allowed reverse transition from impairment to intact cognition. 
  
Semi-competing risks data have also been analyzed using other approaches such as the win ratios and event-specific win ratios  \citep{yang:etal:2022}. 
In the causal inference context such data have recently been considered under the mediation setup with the non-terminal event as a mediator 
  \citep{huang2021causal, xu2022bayesian}.  
  Our multi-state structural models instead consider the total effect of the exposure on all three outcomes: non-terminal event, and terminal event with and without non-terminal event. 
  In addition, since $T_1^a$ may be  $\infty$,
  principal stratification has been considered  when the differences between the (restricted) expectations of $T_1^a$ are estimands of interest \citep{comment2019survivor, nevo2020causal}. 
In contrast we have considered counterfactual systems that are characterized by their respective transition rates, and further developed risk estimates that can be of use in practice. 

For future work, since the IPW estimator is biased if the propensity score model is misspecified, an augmented IPW (AIPW) estimator with doubly robust properties can protect against such model misspecification.
It would also allow us to apply  machine learning or nonparametric methods to 
the propensity score model. \citet{rava2021survival} and \citet{tchetgen2012parametrization} have already developed the AIPW estimator for the marginal structural Cox model, and it is nature to extend their work for the  models in this paper. This is currently under investigation. 
Another future direction is to develop 
sensitivity analysis approaches  for various assumptions including   unmeasured confounding as well as modeling assumptions that are used. 

The R codes developed in this work have been implemented in the R package `semicmprskcoxmsm' that is publicly available on CRAN.

\section*{Acknowledgement}
This research was partially supported by  NIH/NIA grant R03 AG062432. 

\newpage
\renewcommand{\baselinestretch}{1.0}
\bibliographystyle{statinmed}
\bibliography{references}

\newpage

\newpage

\begin{table}[!h]
\centering
\footnotesize
\caption{Simulation results 
with  n = 250; $\beta_1= \beta_2 = 1$ and $\beta_3 = 0.5$. The true value for $\Lambda_{01}$(1) = $\Lambda_{02}$(1) = 1.264, and $\Lambda_{03}$(1) = 2.528. 
} \label{simulation_1}
\begin{tabular}{llc cccc l cccc}
 \cline{1-12} 
                       & \multicolumn{5}{c}{Illness-death (I-D) MSM}                                                                      &  &  \multicolumn{5}{c}{Frailty-based I-D MSM}                       \\ \cline{2-6} \cline{8-12} 
$\sigma^2$     & Par                           & Estimate   & SD      &  model/boot SE                  & model/boot CP                             &  &  Par                         & Estimate & SD     & SE       & CP       \\  \cline{1-12}   
0                     & $\beta_1$                 & 0.995        & 0.211  & 0.214 / 0.219   &   95.6\% / 95.4\%        &  & $\beta_1$                 & 1.063      & 0.197 & 0.201 & 94.8\% \\
                       & $\beta_2$                 & 1.005        & 0.206  & 0.203 / 0.210   &   94.8\% / 95.1\%        &  & $\beta_2$                 & 1.042      & 0.201 & 0.203 & 94.8\% \\
                       & $\beta_3$                 & 0.503        & 0.268  & 0.263 / 0.260   &   94.6\% / 94.8\%        &  & $\beta_3$                 & 0.497      & 0.213 & 0.211 & 95.3\% \\
                       & $\Lambda_{01}$(1)  & 1.219        & 0.264  & 0.259               &   94.8\%                       &  & $\Lambda_{01}$(1)  & 1.323     & 0.275 & 0.280  & 94.6\% \\
                       & $\Lambda_{02}$(1)  & 1.206        & 0.285  & 0.281               &   94.8\%                       &  & $\Lambda_{02}$(1)  & 1.315     & 0.293 & 0.289  & 95.3\% \\
                       & $\Lambda_{03}$(1)  & 2.470        & 0.484  & 0.491               &   96.1\%                       &  & $\Lambda_{03}$(1)  & 2.472     & 0.367 & 0.365  & 95.6\% \\
                       &                &              &         &                     &                                  &  & $\sigma^2$              & 0.038     &  0.018 & 0.030  & 98.0\% \\  \cline{1-12} 
0.5                  & $\beta_1$                 & 0.778        & 0.198  & 0.196 / 0.199   &   80.9\% / 81.3\%         &  & $\beta_1$                 & 1.011      & 0.258 & 0.267 & 96.1\% \\
                       & $\beta_2$                 & 0.782        & 0.204  & 0.209 / 0.204   &   82.4\% / 81.8\%         &  & $\beta_2$                 & 1.005      & 0.261 & 0.267 & 96.1\% \\
                       & $\beta_3$                 & 0.215        & 0.218  & 0.218 / 0.213   &   79.6\% / 78.9\%         &  & $\beta_3$                 &  0.509     & 0.269 & 0.275 & 94.9\% \\
                       & $\Lambda_{01}$(1)  & 1.096        & 0.168  & 0.166               &   77.7\%                        &  & $\Lambda_{01}$(1)  & 1.292      & 0.367 & 0.364 & 94.9\%  \\
                       & $\Lambda_{02}$(1)  & 1.036        & 0.193  & 0.200               &   78.3\%                        &  & $\Lambda_{02}$(1)  & 1.315      & 0.362 & 0.368 & 95.1\%  \\
                       & $\Lambda_{03}$(1)  & 2.749        & 0.406  & 0.403               &   83.5\%                        &  & $\Lambda_{03}$(1)  & 2.460      & 0.518 & 0.521 & 95.6\%  \\
                       &                &              &         &                     &                                  &  &  $\sigma^2$              & 0.572      & 0.199 & 0.193 & 92.9\% \\  \cline{1-12}
1                     & $\beta_1$                & 0.670        & 0.210  & 0.202 / 0.205    &   66.2\% / 65.9\%          &  & $\beta_1$                & 0.993      & 0.258 & 0.270 & 95.5\% \\
                       & $\beta_2$                & 0.679        & 0.198  & 0.201 / 0.195    &   68.6\% / 69.1\%          &  & $\beta_2$                & 0.992      & 0.272 & 0.262 & 94.9\% \\
                       & $\beta_3$                 & 0.104        & 0.243  & 0.239 / 0.240   &   60.0\% / 60.4\%          &  & $\beta_3$                & 0.492      & 0.316 & 0.309 & 93.8\% \\
                       & $\Lambda_{01}$(1)  & 0.984        & 0.172  & 0.177               &   69.1\%                        &  & $\Lambda_{01}$(1)  & 1.290     & 0.395 & 0.394 & 94.5\% \\
                       & $\Lambda_{02}$(1)  & 0.987        & 0.147  & 0.145               &   67.5\%                        &  & $\Lambda_{02}$(1)  & 1.295     & 0.396 & 0.402 & 94.1\% \\
                       & $\Lambda_{03}$(1)  & 3.010        & 0.548  & 0.549               &   71.8\%                        &  & $\Lambda_{03}$(1)  & 2.459     & 0.603 & 0.595 & 95.7\% \\
                       &               &              &         &                     &                                   &  & $\sigma^2$               & 1.089    & 0.270 & 0.275 & 93.6\%    \\   \cline{1-12}
2                     & $\beta_1$                 & 0.561        & 0.201  & 0.205 / 0.202   &   41.8\% / 41.7\%          &  & $\beta_1$                 & 0.985     & 0.301 & 0.291 & 95.9\% \\
                       & $\beta_2$                 & 0.555        & 0.209  & 0.202 / 0.211   &   40.4\% / 39.6\%          &  & $\beta_2$                 & 0.989     & 0.303 & 0.295 & 95.7\% \\
                       & $\beta_3$                 & 0.003        & 0.233  & 0.226 / 0.229   &   33.2\% / 34.0\%          &  & $\beta_3$                 & 0.488     & 0.368 & 0.359 & 94.8\% \\
                       & $\Lambda_{01}$(1)  & 0.920        & 0.134   & 0.128               &   19.4\%                        &  & $\Lambda_{01}$(1)  & 1.233     & 0.330 & 0.333 & 94.3\% \\
                       & $\Lambda_{02}$(1)  & 0.923        & 0.146   & 0.151               &   21.8\%                        &  & $\Lambda_{02}$(1)  & 1.246     & 0.329 & 0.335 & 93.8\% \\
                       & $\Lambda_{03}$(1)  & 3.785        & 0.615   & 0.610               &   11.5\%                         &  & $\Lambda_{03}$(1)  & 2.513     & 0.583 & 0.590 & 96.6\% \\
                       &                &              &         &                      &                                   &  & $\sigma^2$               & 1.912     & 0.318 & 0.326 & 93.1\% \\  \cline{1-12}
\end{tabular}
\end{table}

\newpage

\begin{table}[!h]
\centering
\footnotesize
\caption{Simulation results with  n = 500;  $\beta_1= \beta_2 = 1$ and $\beta_3 = 0.5$. The true value for $\Lambda_{01}$(1) = $\Lambda_{02}$(1) = 1.264, and $\Lambda_{03}$(1) = 2.528. 
} \label{simulation_2}
\begin{tabular}{llc cccc l cccc}
\cline{1-12} 
                       & \multicolumn{5}{c}{Illness-death (I-D) MSM}                                                                      &  &  \multicolumn{5}{c}{Frailty-based I-D MSM}                       \\ \cline{2-6} \cline{8-12} 
$\sigma^2$     & Par                           & Estimate   & SD      & model/boot SE                  &  model/boot CP                            &  &  Par                         & Estimate & SD     & SE       & CP       \\ \cline{1-12} 
0                     & $\beta_1$                 & 1.003        & 0.147  & 0.147 / 0.146   &   95.0\% / 96.0\%        &  &  $\beta_1$               & 1.031      & 0.147 & 0.146  & 94.2\% \\
                       & $\beta_2$                 & 1.000        & 0.141  & 0.137 / 0.145   &   94.8\% / 95.5\%        &  &  $\beta_2$               & 1.040      & 0.145 & 0.147  & 95.5\% \\
                       & $\beta_3$                 & 0.499        & 0.149  & 0.153 / 0.151   &   94.6\% / 95.2\%        &  &  $\beta_3$               & 0.542      & 0.157 & 0.161  & 95.5\% \\
                       & $\Lambda_{01}$(1)  & 1.233        & 0.210  & 0.202               &   95.4\%                       &  & $\Lambda_{01}$(1) & 1.226      & 0.200 & 0.194  & 94.1\% \\
                       & $\Lambda_{02}$(1)  & 1.254        & 0.204  & 0.198               &   94.8\%                       &  & $\Lambda_{02}$(1) & 1.214      & 0.232 & 0.202  & 93.9\% \\
                       & $\Lambda_{03}$(1)  & 2.465        & 0.344  & 0.336               &   94.5\%                       &  & $\Lambda_{03}$(1) & 2.544      & 0.331 & 0.339  & 94.4\% \\
                       &                &              &         &                     &                                  &  & $\sigma^2$             & 0.029      & 0.011 & 0.023  & 98.6\% \\  \cline{1-12}
0.5                  & $\beta_1$                 & 0.762        & 0.141  & 0.143 / 0.141   &   71.2\% / 70.0\%         &  & $\beta_1$                & 1.006      & 0.227 & 0.230 & 95.8\% \\
                       & $\beta_2$                 & 0.775        & 0.151  & 0.148 / 0.146   &   75.4\% / 73.9\%         &  & $\beta_2$                & 0.997      & 0.229 & 0.233 & 96.1\% \\
                       & $\beta_3$                 & 0.219        & 0.158  & 0.160 / 0.158   &   68.0\% / 66.8\%         &  & $\beta_3$                & 0.496      & 0.211 & 0.202 & 94.4\% \\
                       & $\Lambda_{01}$(1)  & 1.183        & 0.138  & 0.130               &   69.4\%                        &  & $\Lambda_{01}$(1) & 1.252      & 0.302 & 0.293 & 94.6\% \\
                       & $\Lambda_{02}$(1)  & 1.178        & 0.146  & 0.139               &   68.6\%                        &  & $\Lambda_{02}$(1) & 1.249      & 0.295 & 0.292 & 94.8\% \\
                       & $\Lambda_{03}$(1)  & 2.734        & 0.361  & 0.356               &   72.1\%                        &  & $\Lambda_{03}$(1) & 2.485      & 0.501 & 0.489 & 95.2\% \\
                       &                &              &         &                     &                                  &  & $\sigma^2$              & 0.566      & 0.179 & 0.186 & 93.3\%  \\  \cline{1-12}
1                     & $\beta_1$                 & 0.667        & 0.146  & 0.137 / 0.143   &   55.2\% / 56.4\%          &  & $\beta_1$                & 1.000     & 0.209 & 0.202 & 94.4\%  \\
                       & $\beta_2$                 & 0.661        & 0.142  & 0.150 / 0.143   &   59.4\% / 56.3\%          &  & $\beta_2$                & 0.998     & 0.211 & 0.202 & 95.2\%   \\
                       & $\beta_3$                 & 0.105        & 0.153  & 0.154 / 0.153   &   47.2\% / 49.4\%          &  & $\beta_3$                & 0.498     & 0.223 & 0.216 & 94.8\%   \\
                       & $\Lambda_{01}$(1)  & 1.018        & 0.124  & 0.123               &   56.7\%                        &  & $\Lambda_{01}$(1) & 1.283     & 0.273 & 0.278 & 96.1\%   \\
                       & $\Lambda_{02}$(1)  & 1.035        & 0.126  & 0.125               &   52.8\%                        &  & $\Lambda_{02}$(1) & 1.289     & 0.269 & 0.275 & 95.5\%    \\
                       & $\Lambda_{03}$(1)  & 2.868        & 0.441  & 0.435               &   62.8\%                        &  & $\Lambda_{03}$(1) & 2.475     & 0.511 & 0.499 & 94.7\%    \\
                       &                &              &         &                     &                                  &  & $\sigma^2$              & 1.063     & 0.189 & 0.184 & 93.9\%   \\  \cline{1-12}
2                     & $\beta_1$                 & 0.563        & 0.149  & 0.142 / 0.144   &   33.8\% / 35.2\%          &  & $\beta_1$                 & 1.009    & 0.268 & 0.273 & 95.6\%   \\
                       & $\beta_2$                 & 0.550        & 0.149  & 0.147 / 0.144   &   34.2\% / 34.4\%          &  & $\beta_2$                 & 1.007    & 0.271 & 0.276 & 95.6\%   \\
                       & $\beta_3$                 & 0.005        & 0.165  & 0.167 / 0.159   &   14.6\% / 13.8\%          &  & $\beta_3$                 & 0.492    & 0.291 & 0.303 & 94.4\%  \\
                       & $\Lambda_{01}$(1)  & 0.920        & 0.104  & 0.099               &   10.8\%                        &  & $\Lambda_{01}$(1)  & 1.244     & 0.302 & 0.300 & 94.6\%   \\
                       & $\Lambda_{02}$(1)  & 0.933        & 0.111   & 0.108               &   12.4\%                        &  & $\Lambda_{02}$(1)  & 1.250    & 0.306 & 0.301 & 95.2\%   \\
                       & $\Lambda_{03}$(1)  & 3.721        & 0.557   & 0.551               &   9.3\%                         &  & $\Lambda_{03}$(1)  &  2.479    & 0.499 & 0.506 & 94.9\%   \\
                       &               &               &         &                     &                                  &  & $\sigma^2$               & 1.924     & 0.255 & 0.252 & 93.0\%   \\  \cline{1-12} 
\end{tabular}
\end{table}

\newpage

\begin{table}[!htp]
\footnotesize
\caption{Event counts  by  heavy versus light alcohol drinking in the HAAS data
} \label{count1}
$$\begin{tabular}{lrrr}
\hline
\multicolumn{1}{l}{}&\multicolumn{1}{c}{\begin{tabular}{c}Heavy Drinking\\ ($n = 426$) \end{tabular}}&\multicolumn{1}{c}{\begin{tabular}{c}Non-Heavy Drinking\\ ($n = 1455$) \end{tabular}}&\multicolumn{1}{c}{\begin{tabular}{c}Overall\\ ($n = 1881$) \end{tabular}}\tabularnewline
\hline
Event&&&\tabularnewline
~~~~censor&$76$ ($17.8\%$)&$281$ ($19.3\%$)&$357$ ($19.0\%$)\tabularnewline
~~~~death without moderate impairment&$135$ ($31.7\%$)&$502$ ($34.2\%$)&$637$ ($34.5\%$)\tabularnewline
~~~~moderate impairment then censor&$50$ ($11.7\%$)&$211$ ($14.5\%$)&$261$ ($13.8\%$)\tabularnewline
~~~~moderate impairment then death&$165$ ($38.7\%$)&$461$ ($31.7\%$)&$626$ ($33.4\%$)\tabularnewline
\hline
\end{tabular}$$ 
\end{table}

\newpage

\begin{table}[!h]
\footnotesize
\caption{
Parameter estimates of heavy ($a=1$) versus non-heavy ($a=0$) drinking using the HAAS data 
} \label{HAAS1}
$$\begin{tabular}{lcccl}
\hline
                                                                       &  Estimate &   SE      &  Hazard Ratio (HR)  &  95\% CI of HR                     \\ \hline
{\bf Illness-death (I-D) MSM} \\                                                                      
moderate impairment                        &    0.212    &  0.082   &     1.237      &   [1.052, 1.453]*      \\
 death without moderate impairment  &    0.237    &  0.102   &     1.268      &  [1.039, 1.547]*      \\
death after moderate impairment      &    0.125    &  0.093    &    1.133      &  [0.944, 1.360]      \\ \hline
{\bf Frailty-based I-D MSM} \\
moderate impairment                        &    0.285    &  0.088   &    1.330       & [1.182, 1.437]*      \\
death without moderate impairment  &    0.317    &  0.109   &    1.373      & [1.083, 1.689]*      \\
death after moderate impairment      &    0.253    &  0.107    &    1.287     & [1.009, 1.538]*      \\ 
$\sigma^2$                                        &    0.881    &  0.113   &         -         & [0.605, 1.054]      \\ \hline
\end{tabular}$$
* indicates statistical significance at $\alpha = 0.05$  two-sided 
\end{table}

\newpage
\begin{table}[!h]
\footnotesize
\caption{Estimated risk difference (RD) and risk ratio (RR) under the illness-death MSM for moderate impairment (MI), death, and death following MI by $t_1 = 8$ years. 
}\label{HAAS2}
$$\begin{tabular}{llrr}
\hline
                       & \multicolumn{1}{c}{Time} & \multicolumn{1}{c}{RD (95\% CI)} & \multicolumn{1}{c}{RR (95\% CI)} \\ \hline
MI      & 5                          & $0.026$ ($0.013$, $0.049$)*      & $1.191$ ($1.075$, $1.408$)*      \\
                       & 10                        & $0.044$ ($0.008$, $0.104$)*      & $1.139$ ($1.026$, $1.283$)*      \\
                       & 15                        & $0.040$ ($0.002$, $0.075$)*      & $1.091$ ($1.001$, $1.237$)*       \\
                       & 20                        & $0.006$ ($-0.030$, $0.029$)     & $1.010$ ($0.938$, $1.111$)       \\ \hline
Death & 5                          & $0.010$ ($-0.007$, $0.012$)      & $1.199$ ($0.995$, $1.358$)      \\
                       & 10                        & $0.029$ ($-0.016$, $0.043$)      & $1.137$ ($0.939$, $1.203$)     \\
                       & 15                        & $0.026$ ($-0.018$, $0.057$)      & $1.081$ ($0.973$, $1.195$)       \\
                       & 20                        & $0.010$ ($-0.041$, $0.048$)      & $1.024$ ($0.912$, $1.057$)       \\ \hline
Death after MI & 10           & $0.030$ ($-0.011$, $0.079$)      & $1.111$ ($0.942$, $1.295$)       \\
                       & 15                        & $0.043$ ($-0.026$, $0.101$)      & $1.056$ ($0.992$, $1.157$)       \\
                       & 20                        & $0.011$ ($-0.018$, $0.028$)      & $1.012$ ($0.986$, $1.116$)       \\ \hline
\end{tabular}$$
 * indicates statistical significance at $\alpha = 0.05$  two-sided.

\end{table}

\newpage

\newpage

%
%
%

\newpage
\section{Supplementary materials}

\subsection{Derivation of $f(t_1,t_2)$, $f_{\infty}(t_2)$ and $S(t,t)$} 
\ba
f_{\infty}(t_{2}) &= \lim_{\Delta \to 0}  \frac{   P\left( T_1 \geq t_2, T_2 \in [t_2,t_2+\Delta)  \right)            }{      \Delta          }  \nonumber \\
&= \lim_{\Delta \to 0}  \frac{   P\left(T_1 \geq t_2, T_2 \in [t_2,t_2+\Delta)  \right)            }{   P\left( T_1 \geq t_2, T_2 \geq t_2  \right)    \Delta          } \times  P\left(T_1 \geq t_2, T_2 \geq t_2  \right)  \nonumber \\
&= \lambda_{2}(t_2)S(t_2,t_2) \nonumber
\end{align} 

We also have:
\ba
f(t_1,t_2) &= \lim_{\Delta \to 0}  \lim_{\delta \to 0}  \frac{   P\left( T_1 \in [t_1,t_1+\delta), T_2 \in [t_2,t_2+\Delta) \right)            }{      \Delta \delta          }  \nonumber \\
& =  \lim_{\Delta \to 0}  \lim_{\delta \to 0}
 P\left(T_1 \geq t_1, T_2 \geq t_1  \right) \times 
 \frac{   P\left(T_1 \in [t_1,t_1+\delta) , T_2 \geq t_1  \right)            }{   P\left( T_1 \geq t_1, T_2 \geq t_1  \right)    \delta          }  \nonumber \\
& \times \frac{P\left( T_2 \in [t_2,t_2+\Delta) \mid T_1 \in [t_1,t_1+\delta), T_2 \ge t_1 \right)}{\Delta}  \nonumber \\
&= \lim_{\Delta \to 0}  \lim_{\delta \to 0} 
P\left(T_1 \geq t_1, T_2 \geq t_1  \right) \times 
\frac{   P\left(T_1 \in [t_1,t_1+\delta) , T_2 \geq t_1  \right)            }{   P\left( T_1 \geq t_1, T_2 \geq t_1  \right)    \delta          }   \nonumber \\
& \times \frac{P\left( T_2 \in [t_2,t_2+\Delta), T_2 \ge t_1 \mid T_1 \in [t_1,t_1+\delta) \right)}{P\left(  T_2 \ge t_1 \mid T_1 \in [t_1,t_1+\delta) \right) \Delta} \nonumber \\
&= \lim_{\Delta \to 0}  \lim_{\delta \to 0} 
P\left(T_1 \geq t_1, T_2 \geq t_1  \right) \times 
\frac{   P\left(T_1 \in [t_1,t_1+\delta) , T_2 \geq t_1  \right)            }{   P\left( T_1 \geq t_1, T_2 \geq t_1  \right)    \delta          }   \nonumber \\
& \times  \frac{   P\left(  T_2 \in [t_2,t_2+\Delta) \mid T_1 \in [t_1,t_1+\delta) \right)          }{   P\left( T_2 \geq t_2 \mid T_1 \in [t_1,t_1+\delta)    \right)    \Delta          } 
\times  \frac{   P\left( T_2 \geq t_2 \mid T_1 \in [t_1,t_1+\delta)    \right)          }
{   P\left( T_2 \geq t_1 \mid T_1 \in [t_1,t_1+\delta)    \right)          } \nonumber \\
&= S(t_1,t_1)\lambda_{1}(t_1)\lambda_{12}(t_2 \mid t_1) 
\exp\left\{ - \int_{t_1}^{t_2} \lambda_{12}(u | t_1) d u    \right\} \nonumber 
\end{align} 


We further have:
\ba
\lambda_1 (t_1 )&= \lim_{\Delta \to 0^+}  \frac{  P\left(T_1 \in [t_1,t_1+\Delta) \mid T_1 \geq t_1, T_2 \geq t_1  \right)            }{      \Delta          } \nonumber \\
&=\frac{ \lim_{\Delta \to 0^+} P\left(T_1 \in [t_1,t_1+\Delta), T_1 \geq t_1, T_2 \geq t_1  \right) /    \Delta } {P(T_1 \geq t_1, T_2 \geq t_1)} \nonumber \\
&=\frac{ \lim_{\Delta \to 0^+} P\left(T_1 \in [t_1,t_1+\Delta), T_2 \geq t_1  \right) /    \Delta } {P(T_1 \geq t_1, T_2 \geq t_1)} \nonumber \\
&=\frac{\int_{t_1}^{+\infty} f(t_1,u) d u } {P(T_1 \geq t_1, T_2 \geq t_1)} \nonumber \\
&=  \frac{- \frac{\partial}{\partial t_1} S(t_1,t_2) \vert_{t_2 = t_1 } }{S(t_1,t_1)}  \nonumber \\
&= - \frac{\partial}{\partial t_1} \log S(t_1,t_2) \vert_{t_2 = t_1 } 
\end{align} 
Similar derivation can be applied to obtain $\lambda_2(t_2)  =
\lim_{\Delta \to 0^+}  {  P\left(T_2 \in [t_2,t_2+\Delta) \mid T_1 \geq t_2, T_2 \geq t_2  \right)            }/{      \Delta          }
=  - \partial \log S(t_1,t_2)/\partial t_2  \vert_{t_1 = t_2 } $. 
By solving the partial derivative equations with the initial condition $S(0,0)=1$, we have 
\ba
S(t,t) = e^{-(\Lambda_1(t) +\Lambda_2(t))}. \nonumber 
\end{align} 
We then 
have \eqref{jointt1} - \eqref{jointt3} in the main text. 

%

\subsection{Causal Identification}

In this section, we prove that our strategy that involves IPW and EM algorithm identifies $\beta$ in the MSM. We consider the frailty-based illness-death MSM. Our weighted EM algorithm which maximizes the logarithm of \eqref{wfull} targets the population maximization problem
\be
\E\left\{ w\log\scaleobj{.7} {\int}   L(  \theta; O \mid b) 
\cdot f(  \theta; b) db  \right\}. \label{popuwfull}
\ee
The goal is to show that the parameters in \eqref{phmm1} - \eqref{phmm3} uniquely maximizes \eqref{popuwfull}. Note that
\begin{align}
    &\E\left\{ w\log\left(\scaleobj{.7} {\int}   L(  \theta; O \mid b) \cdot f(  \theta; b) db\right)  \right\}\\
    & 
    = \int w\log\left(\scaleobj{.7} {\int}   L(  \theta; o \mid b) \cdot f(  \theta; b) db\right)  dF(x_1, x_2, \delta_1, \delta_2, z, a, b)\\
    &\stackrel{\text{consistency}}{=}\int w\log\left(\scaleobj{.7} {\int}   L(  \theta; o^a \mid b) \cdot f(  \theta; b) db\right)  dF(x_1^a, x_2^a, \delta_1^a, \delta_2^a, z, a, b)\\
    &\stackrel{\text{factorization}}{=}\int w\log\left(\scaleobj{.7} {\int}   L(  \theta; o^a \mid b) \cdot f(  \theta; b) db\right)  dF(x_1^a, x_2^a, \delta_1^a, \delta_2^a|a, z, b)F(a|z)F(z)F(b)\\
    &\stackrel{\text{exchangeability}}{=}\int w\log\left(\scaleobj{.7} {\int}   L(  \theta; o^a \mid b) \cdot f(  \theta; b) db\right)  dF(x_1^a, x_2^a, \delta_1^a, \delta_2^a|z, b)F(a|z)F(z)F(b)\\
    &\stackrel{\text{weights definition}}{=}\sum_{a = 0, 1}\int\log\left(\scaleobj{.7} {\int}   L(  \theta; o^a \mid b) \cdot f(  \theta; b) db\right)  dF(x_1^a, x_2^a, \delta_1^a, \delta_2^a|z, b)F(z)F(b)\\
    &=\sum_{a = 0, 1}\int\log\left(\scaleobj{.7} {\int}   L(  \theta; o^a \mid b) \cdot f(  \theta; b) db\right)  dF(x_1^a, x_2^a, \delta_1^a, \delta_2^a, z, b)\\
    &=\sum_{a = 0, 1}\int\log\left(\scaleobj{.7} {\int}   L(  \theta; o^a \mid b) \cdot f(  \theta; b) db\right)  d\int L(  \theta_0; o^a \mid b) \cdot f(  \theta_0; b)db,
\end{align}
where we denote $\theta_0$ as the truth. Since the logarithm function is concave, by Jensen's inequality, this maximization problem achieves its maximum at $\theta_0$. Generally speaking, this proves that whatever techniques that work under a randomized experiment, in our case, the Markov illness-death model and its associated estimation procedure, shall continue to work as long as  these weights are properly leveraged, for non-randomized yet exchangeable treatment assignment.

\subsection{Variance-covariance  under the illness-death MSM} \label{usual_var}

For the $i$th individual, let the at-risk process for non-terminal event, terminal event without non-terminal event, and terminal event following 
 non-terminal event as 
$Y_{1i}(t) =I(X_{1i} \geq t)$, $Y_{2i}(t) = I(X_{2i} \geq t, X_{1i} \geq t )$, 
and $Y_{3i}(t) = I(X_{2i} \geq t \geq X_{1i})$. 
It is also convenient to introduce the following notation: 
\begin{align*}
S^{(1)}_{1w} ( \hat{\beta}_1 ; t ) =& \sum_{\ell=1}^n w_\ell Y_{1\ell} ( t ) A_\ell \exp(\hat{\beta}_1 A_\ell ), \  S^{(0)}_{1w} ( \hat{\beta}_1 ; t ) = \sum_{\ell=1}^n w_\ell Y_{1\ell} ( t ) \exp(\hat{\beta}_1 A_\ell );\\ 
S^{(1)}_{2w} ( \hat{\beta}_2 ; t ) =& \sum_{\ell=1}^n w_\ell Y_{2\ell} ( t ) A_\ell \exp(\hat{\beta}_2 A_\ell ), \ S^{(0)}_{2w} ( \hat{\beta}_2 ; t ) = \sum_{\ell=1}^n w_\ell Y_{2\ell} ( t )  \exp(\hat{\beta}_2 A_\ell ); \\ 
S^{(1)}_{3w} ( \hat{\beta}_3 ; t ) =& \sum_{\ell=1}^n w_\ell Y_{3\ell} ( t ) A_\ell \exp(\hat{\beta}_3 A_\ell ), \ S^{(0)}_{3w} ( \hat{\beta}_3 ; t ) = \sum_{\ell=1}^n w_\ell Y_{3\ell} ( t )  \exp(\hat{\beta}_3 A_\ell ).
\end{align*}
Then the robust sandwich variance estimator 
 is given by 
$
V(  \hat{\beta}  ) = B(  \hat{\beta}  ) M( \hat{\beta} ) B(  \hat{\beta}  ),
$
where 
$B( \hat{\beta} )  =  - 
\partial^2 \log L_w(\beta) / \partial \beta^2  \vert_{\beta = \hat{\beta} } /n
= [b_{jj}]_{j=1,2,3}$   is a diagonal matrix, 
\ba
b_{11}  =& - \frac{1}{n} \sum_{i=1}^{n} w_i \delta_{i1} \left\{  A_i -  \frac {   S^{(1)}_{1w} ( \hat{\beta}_1 ; X_{1i}  ) } {  S^{(0)}_{1w} ( \hat{\beta}_1 ; X_{1i}  ) }    \right\}, \nonumber \\ 
b_{22}  =& - \frac{1}{n} \sum_{i=1}^{n} w_i (1-\delta_{i1})\delta_{i2} \left\{  A_i -    \frac { S^{(1)}_{2w} ( \hat{\beta}_2 ; X_{1i}  ) } {  S^{(0)}_{2w} ( \hat{\beta}_2 ; X_{1i}  ) }    \right\}, \nonumber \\ 
b_{33}  =& - \frac{1}{n} \sum_{i=1}^{n} w_i \delta_{i1}\delta_{i2} \left\{  A_i -     \frac {  S^{(1)}_{3w} ( \hat{\beta}_3 ; X_{2i}  ) } {  S^{(0)}_{3w} ( \hat{\beta}_3 ; X_{2i}  )}    \right\}; \nonumber
\end{align}
and  $M(  \hat{\beta}  ) =  
\sum_{i=1}^n  \hat{U}^{(i)}(\hat{\beta} ) \hat{U}^{(i)}( \hat{\beta} ) ^ {'} /n $ with 
\ba
U_1^{(i)}(\hat{\beta}_1)= & w_i \delta_{1i} \left\{ A_i -  \frac {   S^{(1)}_{1w} ( \hat{\beta}_1 ; X_{1i}  ) } {  S^{(0)}_{1w}( \hat{\beta}_1 ; X_{1i}  ) } \right\} \nonumber \\
& - w_{i} \cdot \sum_{\ell=1}^{n} w_\ell \delta_{1\ell} \frac{ Y_{1i}(X_{1\ell}) \exp(\beta_1 A_i)} { S^{(0)}_{1w}(\hat{\beta}_{1};X_{1 \ell}) }  \left\{ A_i - \frac{S^{(1)}_{1w}(\hat{\beta}_1;X_{1 \ell })}{S^{(0)}_{1w}(\hat{\beta}_1;X_{1 \ell})}  \right\},  \nonumber \\
U_2^{(i)}(\hat{\beta}_2)= & w_i (1-\delta_{1i})\delta_{2i} \left\{ A_i -  \frac {   S^{(1)}_{2w} ( \hat{\beta}_2 ; X_{1i}  ) } {  S^{(0)}_{2w}( \hat{\beta}_2 ; X_{1i}  ) } \right\} \nonumber \\
& - w_{i} \cdot \sum_{\ell=1}^{n} w_\ell (1-\delta_{1\ell})\delta_{2\ell} \frac{ Y_{2i}(X_{1\ell}) \exp(\beta_2 A_i)} { S^{(0)}_{2w}(\hat{\beta}_{2};X_{1 \ell}) }  \left\{ A_i - \frac{S^{(1)}_{2w}(\hat{\beta}_2;X_{1 \ell })}{S^{(0)}_{2w}(\hat{\beta}_2;X_{1 \ell})}  \right\},  \nonumber \\
U_3^{(i)}(\hat{\beta}_3)= & w_i \delta_{1i}\delta_{2i} \left\{ A_i -  \frac {   S^{(1)}_{3w} ( \hat{\beta}_3 ; X_{2i}  ) } {  S^{(0)}_{3w} ( \hat{\beta}_3 ; X_{2i}  ) } \right\} \nonumber \\
& - w_{i} \cdot \sum_{\ell=1}^{n} w_\ell \delta_{1\ell}\delta_{2\ell} \frac{ Y_{3i}(X_{2\ell}) \exp(\beta_3 A_i)} { S^{(0)}_{3w}(\hat{\beta}_{3};X_{2 \ell}) }  \left\{ A_i - \frac{S^{(1)}_{3w}(\hat{\beta}_3;X_{2 \ell })}{S^{(0)}_{3w}(\hat{\beta}_3;X_{2 \ell})}  \right\}. \nonumber 
\end{align}


\newpage
\subsection{Proof of Lemma \ref{lemma1}} 
\begin{proof}
From (\ref{wfull}) in the main text, we have: 
\begin{align}
l_{w}(  \theta; O) = & \log L_{w}(  \theta; O)  \nonumber\\
=& \log \bigg\{ \scaleobj{.8}{ \prod_{i} } \big( \scaleobj{.7} {\int}  L(  \theta; O_{i} \mid b_{i}) \cdot f(\theta;b_{i}) db_{i}  \big)^{w_{i}} \bigg\}  \nonumber\\
=& \sum_{i} w_{i} \log \int \frac{L(  \theta; O_{i} \mid b_{i}) \cdot  f(\theta; b_{i}) }{f(b_{i}|  \theta^{(k)},O_{i})} f(b_{i}|  \theta^{(k)},O_{i}) d b_{i} 
 \nonumber\\
=& \sum_{i} w_{i}  \log \E
\bigg[ \left. \frac{L(  \theta; O_{i} \mid b_{i}) \cdot  f(\theta;b_{i}) }{f(b_{i}|  \theta^{(k)},0_{i})} \right | 
 \theta^{(k)}, O_{i} \bigg] \label{line4} \\
\ge& \sum_{i} w_{i} \E
\bigg[ \log \big(  \frac{L(  \theta; O_{i} \mid b_{i})\cdot  f(\theta;b_{i})  }{f(b_{i}|  \theta^{(k)},O_{i})} \big)  \mid   \theta^{(k)},O_{i}\bigg]  \label{line5} \\
=& \sum_{i} \E_{  \theta^{(k)}} \big[ w_{i} \cdot l(  \theta; O_{i} \mid b_{i}) \mid O_{i} \big] + \E \big[ w_{i} \cdot \log f(  b_{i}; \theta)) \mid   \theta^{(k)},O_{i} \big] \nonumber\\
& - \E \big[ w_{i} \cdot  \log f(b_{i} \mid   \theta^{(k)},O_{i}) \mid   \theta^{(k)},O_{i} \big]  \nonumber\\
= & Q(  \theta;   \theta^{(k)}) - \sum_{i} w_{i}  \E \big[ \log f(b_{i} \mid   \theta^{(k)},O_{i}) \mid   \theta^{(k)},O_{i} \big], \nonumber
\end{align}
where the inequality above comes from Jensen's inequality. 
If $  \theta =    \theta^{(k)}$, \eqref{line4} becomes 
\begin{align*}
\sum_{i} w_{i}  \log \E \bigg[ \frac{L(  \theta^{(k)}; O_{i} \mid b_{i}) \cdot  f(  \theta^{(k)};b_{i}) }{f(b_{i}|  \theta^{(k)},O_{i})} \mid   \theta^{(k)},O_{i}\bigg]
 =& \sum_{i} w_{i}  \log \E \bigg[f(O_{i} \mid    \theta^{(k)}) \mid   \theta^{(k)}, O_{i} \bigg] \\
 =& \sum_{i} w_{i}  \log  f(O_{i} \mid    \theta^{(k)}) \\
 =& \sum_{i} w_{i}   \E \bigg[ \log f(O_{i} \mid    \theta^{(k)}) \mid   \theta^{(k)}, O_{i} \bigg],
\end{align*}
which equals \eqref{line5}. \\
Then we have $l_{w}(  \theta^{(k)}; O) = Q(  \theta^{(k)};   \theta^{(k)}) - \sum_{i} w_{i}  \E \big[ \log f(b_{i} \mid   \theta^{(k)},O_{i}) \mid   \theta^{(k)},O_{i} \big] $.
Therefore 
\begin{align*}
& l_{w}(  \theta^{(k+1)}; O) - l_{w}(  \theta^{(k)}; O) \\
\ge & Q(  \theta^{(k+1)};  \theta^{(k)}) - Q(  \theta^{(k)};  \theta^{(k)}) - \bigg(\sum_{i} w_{i}  \E \big[ \log f(b_{i} \mid   \theta^{(k)},O_{i}) \mid   \theta^{(k)},O_{i} \big]  \\
& - \sum_{i} w_{i}  \E \big[ \log f(b_{i} \mid   \theta^{(k)},O_{i}) \mid   \theta^{(k)},O_{i} \big]  \bigg) \\
= & Q(  \theta^{(k+1)};  \theta^{(k)}) - Q(  \theta^{(k)};  \theta^{(k)}). 
\end{align*}
Since $  \theta^{(k+1)}$ maximizes  $Q(  \theta,  \theta^{(k)})$, 
$Q(  \theta^{(k+1)};  \theta^{(k)}) - Q(  \theta^{(k)};  \theta^{(k)}) \ge 0$. 
Therefore $l_w(  \theta^{(k+1)};O) \ge l_w(  \theta^{(k)};O)$, 
and $L_w(  \theta^{(k+1)};O) \ge L_w(  \theta^{(k)};O)$. 
\end{proof}

\newpage
\subsection{Detailed calculation of $E(h(b_{i})|O_i,\tilde{  \theta})$} 
We have 
\begin{align*}
E(h(b_{i})|O_i;\tilde{ \theta}) &= \int h(b_{i}) \cdot  f(b_{i} \mid O_i;\tilde{ \theta}) d b_{i} \\
&= \int h(b_{i}) \cdot  \frac{f(O_i, b_{i}; \tilde{  \theta})}{f(O_i;\tilde{  \theta})} d b_{i} \\
&= \int  h(b_{i}) \cdot  \frac{f(O_i \mid b_{i};\tilde{  \theta})  f(b_{i};\tilde{  \theta})}{f(O_i;\tilde{  \theta})} d b_{i},
\end{align*}
where
\begin{align*}
f(O_i;\tilde{  \theta}) &= \int f(O_i, b_{i}; \tilde{  \theta}) d b_{i} \\
&= \int f(O_i \mid b_{i};\tilde{  \theta}) \cdot  f(b_{i};\tilde{  \theta}) d b_{i}.
\end{align*}
After plugging in model based quantities, we have
\begin{align*}
f(O_i;\tilde{  \theta}) &= \int \bigg[ \big\{  \tilde{\lambda}_{01}\left(X_{1i}\right) \exp{(\tilde{\beta}_{1}A_{i}+b_{i}) \big\}}^{\delta_{1i}} 
\exp\{- \tilde{\Lambda}_{01}(X_{1i}) \exp(\tilde{\beta}_{1}A_{i}+b_{i}) \} \\
&\cdot    \big\{   \tilde{\lambda}_{02}\left(X_{2i}\right) \exp{(\tilde{\beta}_{2}A_{i}+b_{i})} \big\}^{\delta_{2i}(1-\delta_{1i})} 
\exp\{-\tilde{\Lambda}_{02}(X_{1i}) \exp(\tilde{\beta}_{2}A_{i}+b_{i})\} \\
&\cdot    \big\{  \tilde{\lambda}_{03}\left(X_{2i}\right) \exp{(\tilde{\beta}_{3}A_{i}+b_{i})} \big\}^{\delta_{2i}\delta_{1i}} 
\exp \big\{  - \tilde{\Lambda}_{03}(X_{1i},X_{2i}) \exp(\tilde{\beta}_{3}A_{i}+b_{i}) \big\} \bigg] \\
&\cdot  \bigg[ \frac{ \exp ( -\frac{b_{i}^2}{2 \tilde{\sigma}^2} ) }{\sqrt{2\pi \tilde{\sigma}^2}} \bigg] d b_{i}.
\end{align*}
Then we have 
\begin{align*}
E(h(b_{i})|O_i;\tilde{ \theta}) &= \int \frac{h(b_{i})}{f(O_i;\tilde{  \theta})} \cdot  \bigg[ \big\{  \tilde{\lambda}_{01}\left(X_{1i}\right) \exp{(\tilde{\beta}_{1}A_{i}+b_{i}) \big\}}^{\delta_{1i}} 
\exp\{- \tilde{\Lambda}_{01}(X_{1i}) \exp(\tilde{\beta}_{1}A_{i}+b_{i}) \} \\
&\cdot   \big\{   \tilde{\lambda}_{02}\left(X_{2i}\right) \exp{(\tilde{\beta}_{2}A_{i}+b_{i})} \big\}^{\delta_{2i}(1-\delta_{1i})} 
\exp\{-\tilde{ \Lambda}_{02}(X_{1i}) \exp(\tilde{\beta}_{2}A_{i}+b_{i})\} \\
&\cdot    \big\{  \tilde{\lambda}_{03}\left(X_{2i}\right) \exp{(\tilde{\beta}_{3}A_{i}+b_{i})} \big\}^{\delta_{2i}\delta_{1i}} 
\exp \big\{  - \tilde{\Lambda}_{03}(X_{1i},X_{2i}) \exp(\tilde{\beta}_{3}A_{i}+b_{i}) \big\} \bigg] \\
&\cdot  \bigg[ \frac{ \exp ( -\frac{b_{i}^2}{2 \tilde{\sigma}^2} ) }{\sqrt{2\pi \tilde{\sigma}^2}} \bigg] d b_{i}.
\end{align*}
Numerical methods such as adaptive Gaussian quadrature can be
 used to calculate the integral, 
 which is what we use in this paper.

\newpage
\subsection{Bayesian bootstrap}

  For each  bootstrap sample: 
\begin{itemize}
  \item Generate $n$ standard exponential (mean and variance 1) random variates : $u_1, u_2,..., u_n$;
  \item The weights for the Bayesian bootstrap are: $w_{i}^{boot} = u_i / \bar{u}$, $i = 1, 2, ..., n$, where $\bar{u} = n^{-1}\sum_{i=1}^{n} u_i$;
  \item Calculate the propensity score and IP weights $w_{i}^{IPW}$ based on Bayesian bootstrap weighted data, and assigned the weights for fitting the MSM general Markov model as $w_i = w_{i}^{boot} * w_{i}^{IPW}$.
  \item After obtaining $\hat{\theta}$ and $\hat{b}_i$, for each individual $i$, calculate the IRR and IRD by plugging $\hat{\theta}, \hat{b}_i$ and $a=0, a=1$ separately into \eqref{generalcifb1} -  \eqref{generalcifb3} from main text at time $t$: $\hat{F_1}_i(t \mid b_i ; 1) - \hat{F_1}_i(t \mid b_i ; 0)$, $\hat{F_2}_i(t \mid b_i ; 1) - \hat{F_2}_i(t \mid b_i ; 0)$ and $\hat{ F_{12} }_i(t_1, t \mid b_i ; 1) - \hat{ F_{12} }_i(t_1, t \mid b_i ; 0)$, etc..
\end{itemize}
The 95\% prediction intervals (PI) are obtained by the normal approximation using bootstrap standard error.

\subsection{Details for the simulation steps}

 Following 
 \cite{jiang2015simulation}, 
  from \eqref{jointt2} in the main text and $\lambda_{01}\left(t \right) = \lambda_{02}\left(t \right) = 2\exp(-t)I(0 \le t \le 3) + 2\exp(-3)I(t \ge 3)$ and $\lambda_{03}(t) = 2\lambda_{01}(t)$, 
  we have 
\ba
P(T_{1} = \infty) &= \int_{0}^{+\infty} f_{\infty}(t \mid b) d t \nonumber \\
&= \int_{0}^{+\infty} e^{\beta_{2}z+b} \lambda_{02}(t) \exp \big\{-e^{\beta_{1}z+b} \Lambda_{01}(t)-e^{\beta_{2}z+b} \Lambda_{02}(t) \big\} d t \nonumber \\
&= \frac{e^{\beta_{2}z}}{e^{\beta_{1}z} + e^{\beta_{2}z}}.  \label{probt1}
\end{align}
We 
can also derive the conditional marginal density of $T_{1}$ when $T_{1} < \infty$ from $f(t_{1},t_{2}\mid b)$ as: 
\ba
f(t_{1} \mid b) &= \int_{t_{1}}^{+\infty} f(t_{1}, t \mid b) d t \nonumber \\
&= \int_{t_{1}}^{+\infty}  e^{\beta_{1}z+\beta_{3}z+2b} \lambda_{01}(t_{1}) \lambda_{03}(t) \exp \big\{-e^{\beta_{1}z+b} \Lambda_{01}(t_{1})-e^{\beta_{2}z+b} \Lambda_{02}(t_{1}) -e^{\beta_{3}z+b} \Lambda_{03}(t_{1},t)\big\} d t \nonumber \\
&= e^{\beta_{1}z+b} \lambda_{01}(t_{1})  \exp \big\{-e^{\beta_{1}z+b} \Lambda_{01}(t_{1})-e^{\beta_{2}z+b} \Lambda_{02}(t_{1}) \big\}  \nonumber \\
&\quad \cdot \int_{t_1}^{\infty}  \exp \big(-e^{\beta_{3}z+b} \Lambda_{03}(t_{1},t)\big) d \big\{  e^{\beta_{3}z+b}  \Lambda_{03}(t) \big\} \nonumber \\ 
&= e^{\beta_{1}z+b} \lambda_{01}(t_{1})  \exp \big\{-e^{\beta_{1}z+b} \Lambda_{01}(t_{1})-e^{\beta_{2}z+b} \Lambda_{02}(t_{1}) \big\} \nonumber \\ 
&=   e^{\beta_{1}z+b} \lambda_{01}(t_{1})  \exp \big\{-e^{\beta_{1}z+b} \Lambda_{01}(t_{1})-e^{\beta_{2}z+b} \Lambda_{01}(t_{1}) \big\}. \label{margt1}
\end{align}
Therefore the conditional survival functions of $T_{1}$ conditional on $b$ are 
\ba
S_{1}(t_{1} \mid b) &= P(t_{1} \le T_{1} < \infty) + P(T_{1} = \infty)  \nonumber \\
&= \int_{t_{1}}^{+\infty} f(t \mid b) d t + Pr(T_{1} = \infty)  \nonumber \\
&= \frac{e^{\beta_{1}z}}{e^{\beta_{1}z} + e^{\beta_{2}z}} \exp \big\{-(e^{\beta_{1}z+b} + e^{\beta_{2}z+b})\Lambda_{01}(t_{1})\big\} +  \frac{e^{\beta_{2}z}}{e^{\beta_{1}z} + e^{\beta_{2}z}},
\end{align}
and 
\ba
S_{1}(t_{1} \mid T_{1}< \infty, b) &= \frac{S_{1}(t_{1},T_{1}< \infty \mid b)}{1-Pr(T_{1}=\infty)}  \nonumber \\
&=  \exp \big\{-(e^{\beta_{1}z+b} + e^{\beta_{2}z+b})\Lambda_{01}(t_{1})\big\}. \label{survt1} 
\end{align}
We also need 
 the conditional joint probability $P(T_{2} > t_{2}, T_{1} \in [t_{1},t_{1}+\Delta t] \mid b)$, 
 $t_{1} < t_{2} < \infty$: 
\ba
&P(T_{2} > t_{2}, T_{1} \in [t_{1},t_{1}+\Delta t] \mid b) 
= \int_{t_{2}}^{+\infty} f(t_{1},t \mid b) d t  \nonumber \\
= & e^{\beta_{1}z+b}\lambda_{01}(t_{1})  
\cdot  \exp \big[-e^{\beta_{1}z+b} \Lambda_{01}(t_{1})-e^{\beta_{2}z+b} \Lambda_{02}(t_{1})-e^{\beta_{3}z+b} (\Lambda_{03}(t_{2})-\Lambda_{03}(t_{1})) \big] \nonumber \\
= &  e^{\beta_{1}z+b} \lambda_{01}(t_{1})  
\cdot \exp \big[-e^{\beta_{1}z+b} \Lambda_{01}(t_{1})-e^{\beta_{2}z+b} \Lambda_{01}(t_{1})-2e^{\beta_{3}z+b} (\Lambda_{01}(t_{2})-\Lambda_{01}(t_{1})) \big].
\end{align}
Therefore, the conditional survival function for $T_{2}$ given $T_{1} = t_{1} < \infty$ and $b$ is:
\ba
S_{21}( t_{2} \mid t_{1}, b) &= P(T_{2}>t_{2} \mid T_{1} = t_{1}, b) = \frac{P(T_{2} > t_{2}, T_{1} \in [t_{1},t_{1}+\Delta t] \mid b)}{f(t_{1} \mid b)}  \nonumber \\
&= \exp \big(-2e^{\beta_{3}z+b} \{ \Lambda_{01}(t_{2})-\Lambda_{01}(t_{1}) \} \big),  \label{survt2t1} 
\end{align}
and the conditional survival function for $T_{2}$ given $T_{1} =  \infty$ and $b$ is
\ba
S_{21}( t_{2} \mid T_{1}=\infty, b) &= P(T_{2}>t_{2} \mid T_{1}=\infty, b) = \frac{P(T_{2} > t_{2}, T_{1}=\infty \mid b)}{Pr(T_{1}=\infty)}  \nonumber \\
&= \frac{\int_{t_{2}}^{+\infty} f_{\infty}(t \mid b) d t }{Pr(T_{1}=\infty)} \nonumber \\
&= \exp \left\{ -(e^{\beta_{1}z+b} + e^{\beta_{2}z+b})\Lambda_{01}(t_{2}) \right\}.   \label{survt2t1} 
\end{align}

Based on the above, 
we can generate the event time $T_{1}, T_{2}$: with probability $P(T_{1}=\infty)$, we can generate $T_{2}$ from $S_{21}( t_{2} \mid T_{1}=\infty, b)$, and with probability $1-P(T_{1}=\infty)$, we can generate $T_{1}$ from $S_{1}(t_{1} \mid T_{1}< \infty, b)$, then generate $T_{2}$ from $S_{21}( t_{2} \mid t_{1}, b)$ conditioning on the observed value of $T_{1} = t_{1}$.

\newpage
\subsection{HAAS data analysis}

\begin{figure}[!h]
\centering 
\includegraphics[width=150mm]{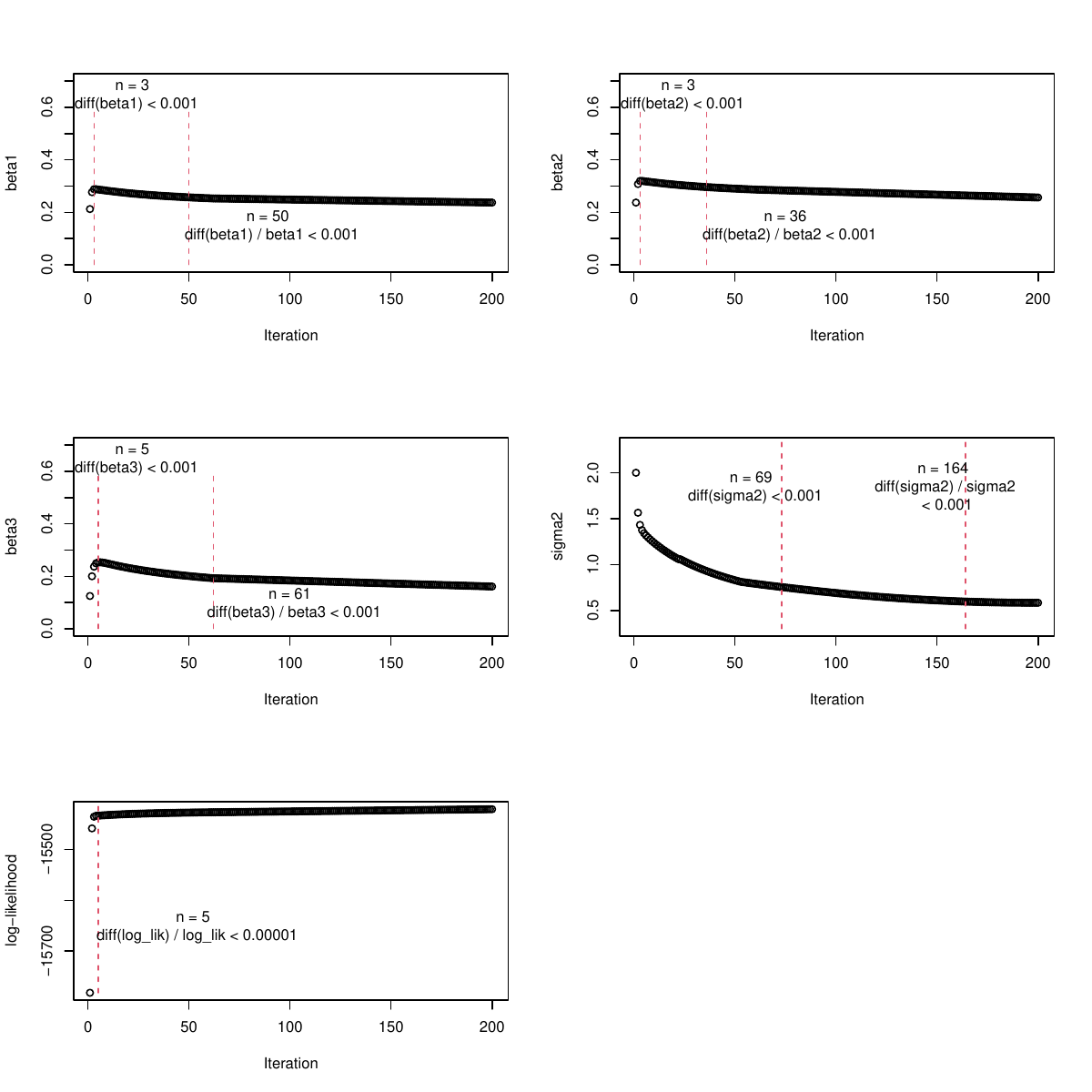}
\caption{Convergence plots for the HAAS data analysis  under the frailty-based illness-death MSM}
\end{figure}


\newpage
\begin{figure}[!h]
\centering 
\includegraphics[width=100mm]{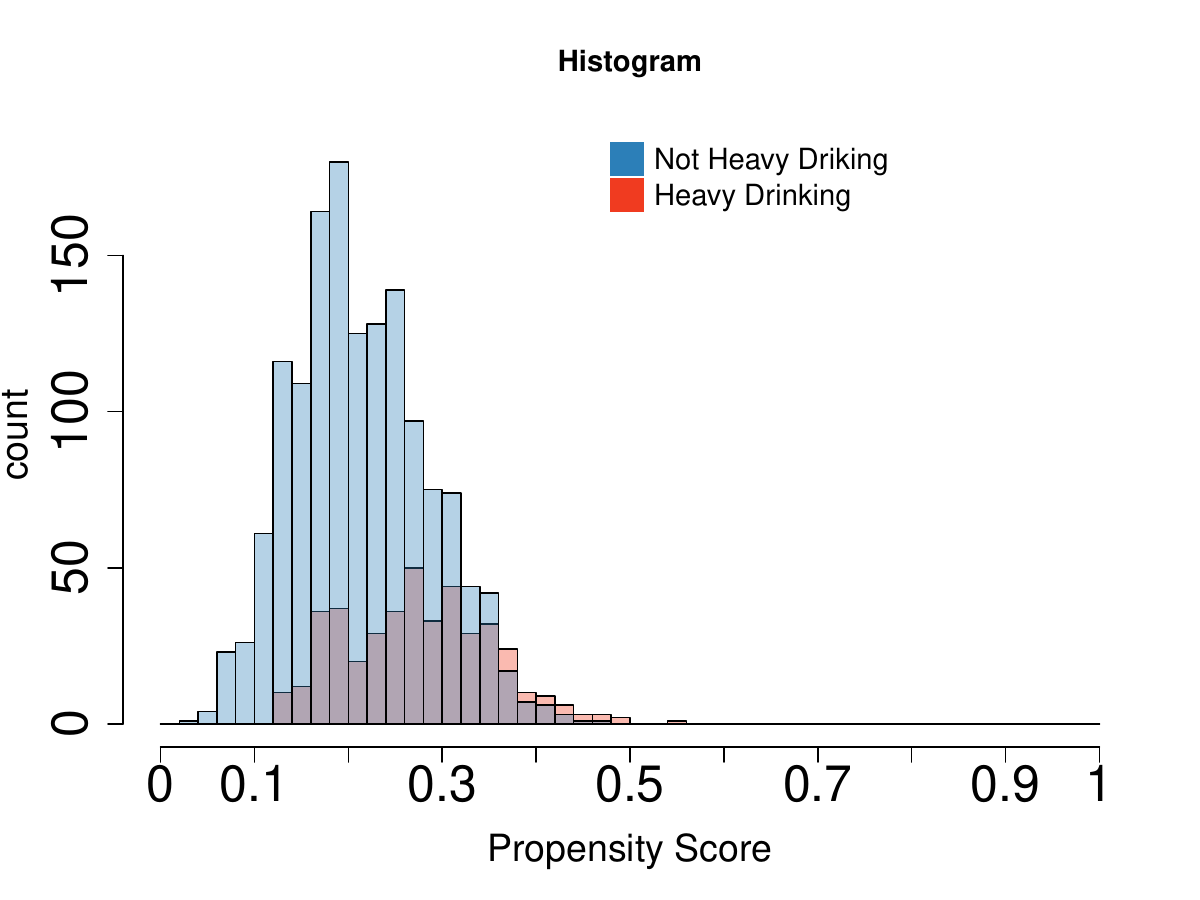}\\
\includegraphics[width=80mm]{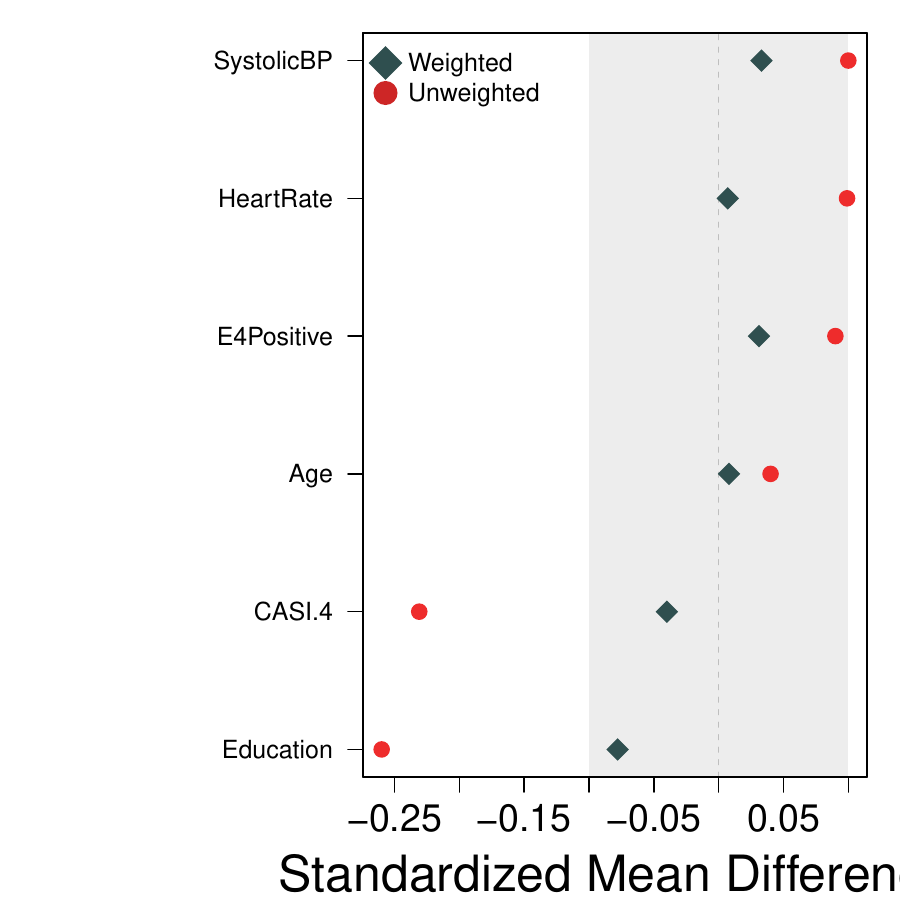}
\caption{Distribution of PS (top) and SMD plot (bottom)} \label{ps_plot}
\end{figure}

\end{document}